\documentclass[12pt, draftclsnofoot, onecolumn]{IEEEtran}
\ifCLASSINFOpdf
\else
\fi
\usepackage{multirow}
\usepackage{color}
\usepackage{graphicx}
\usepackage[square,comma,sort&compress]{natbib}
\usepackage{amsmath}
\usepackage{amssymb}
\DeclareMathSizes{10}{9}{7}{6}
\IEEEaftertitletext{\vspace{-2\baselineskip}}
\begin{document}
\title{Binary Compressive Sensing via Analog Fountain Coding}
\author{Mahyar~Shirvanimoghaddam,~\IEEEmembership{Member,~IEEE,}
                          Yonghui~Li,~\IEEEmembership{Senior~Member,~IEEE,}
                       Branka~Vucetic,~\IEEEmembership{Fellow,~IEEE,}
                       and~Jinhong~Yuan,~\IEEEmembership{Senior~Member,~IEEE,}
\thanks{The material in this paper was presented in part at the 2014 IEEE Global Communications Conference, Austin, TX.

Mahyar Shirvanimoghaddam, Yonghui Li and Branka vucetic are with the Center of Excellence in Telecommunications,
School of Electrical and Information Engineering, The University of Sydney,
Sydney, NSW 2006, Australia (e-mail: mahyar.shirvanimoghaddam@sydney.edu.au; yonghui.li@sydney.edu.au; branka.vucetic@sydney.edu.au).

Jinhong Yuan is with the School of Electrical Engineering and Telecommunications, The University of New South Wales, Sydney, NSW 2052, Australia (e-mail:j.yuan@unsw.edu.au).}}
\maketitle
\begin{abstract}
In this paper, a compressive sensing (CS) approach is proposed for sparse binary signals' compression and reconstruction based on analog fountain codes (AFCs). In the proposed scheme, referred to as the analog fountain compressive sensing (AFCS), each measurement is generated from a linear combination of $L$ randomly selected binary signal elements with real weight coefficients. The weight coefficients are chosen from a finite weight set and $L$, called measurement degree, is obtained based on a predefined degree distribution function. We propose a simple verification based reconstruction algorithm for the AFCS in the noiseless case. The proposed verification based decoder is analyzed through SUM-OR tree analytical approach and an optimization problem is formulated to find the optimum measurement degree to minimize the number of measurements required for the reconstruction of binary sparse signals. We show that in the AFCS, the number of required measurements is of $\mathcal{O}(-n\log(1-k/n))$, where $n$ is the signal length and $k$ is the signal sparsity level. We then consider the signal reconstruction of AFCS in the presence of additive white Gaussian noise (AWGN) and the standard message passing decoder is then used for the signal recovery. Simulation results show that the AFCS can perfectly recover all non-zero elements of the sparse binary signal with a significantly reduced number of measurements, compared to the conventional binary CS and $\ell_1$-minimization approaches in a wide range of signal to noise ratios (SNRs). Finally, we show a practical application of the AFCS for the sparse event detection in wireless sensor networks (WSNs), where the sensors' readings can be treated as measurements from the CS point of view.
\end{abstract}
\IEEEpeerreviewmaketitle
\section{Introduction}
\IEEEPARstart{C}{ompressive sensing} (CS) has drawn a lot of interests in recent years as a revolutionary
signal sampling paradigm. The main idea behind CS is the possibility to reconstruct a sparse signal from an under-determined system of linear equations through a computationally efficient convex programming algorithm. Let us consider the reconstruction problem of a sparse signal  $\textbf{x}\in \mathbb{R}^n$ from the linear measurements $\textbf{y}\in\mathbb{R}^m$, where $m<n$, $\textbf{y}=\textbf{Gx}$, and $\textbf{G}\in\mathbb{R}^{m\times n}$ is the measurement matrix, modeling the measurement (sensing) process. We assume that the signal $\textbf{x}$ is  $k$-sparse, which means that exactly $k\ll n$ elements of $\textbf{x}$ are nonzero \cite{NosiyCSCoding}, but we do not know the locations of the nonzero entries of \textbf{x}. Due to the sparsity of \textbf{x}, one could try to compute \textbf{x} by solving the following optimization problem:
\begin{align}
\label{NPhard0}
\min_{\textbf{x}}||\textbf{x}||_0~~~~\text{s.t.}~~~~\textbf{Gx}=\textbf{y}.
\end{align}
However, solving (\ref{NPhard0}) is an NP-hard problem and thus practically not feasible \cite{CSDonoho}. Instead, a convex relaxation problem can be formulated as follows, which can be solved efficiently via linear or quadratic programming techniques.
\begin{align}
\label{Relax1s}
\min_{\textbf{x}}||\textbf{x}||_1~~~~\text{s.t.}~~~~\textbf{Gx}=\textbf{y}.
\end{align}
As shown in \cite{MesureSurvey,robustUncer,NearOptimal,CSDonoho}, under certain conditions on the matrix \textbf{G} and the sparsity of \textbf{x}, both (\ref{NPhard0}) and (\ref{Relax1s}) have the same unique solution. The Restricted Isometry Property (RIP) and the coherence of matrix \textbf{G} are to date the most widely used tools to derive such conditions. A comprehensive introduction to compressive sensing can be found in excellent review papers \cite{CSBaraniuk,IntroCSCandes}.

Unlike the conventional CS theory which considers only the sparsity of the signal, in \cite{baraniuk2010model} a more realistic signal model and CS approach, referred to as the Model-based CS, was introduced by considering some structure among the signal coefficients. Model-based CS and its variations has been widely studied in the literature \cite{baraniuk2010model,cohen2009compressed,eldar2009robust,ganesh2009separation,stojnic2009reconstruction,tropp2006algorithms,zelinski2008sparsity}. Due to the known structure of the sparse signal in model-based CS, signal models provide two major advantages to CS. First, the number of measurements required for the reconstruction of the sparse signal is significantly reduced. Second, a more accurate and robust recovery is achieved as we can better differentiate true signal information from recovery artifacts \cite{baraniuk2010model}. In this paper, we are particularly interested in the reconstruction of sparse binary signals using linear measurements. Binary CS is a special case of model based CS which has many applications, such as event detection in wireless sensor networks (WSNs), group testing, and spectrum sensing for cognitive radios. More practical settings and similar discrete types of compressive sensing scenarios can be found in \cite{UniqueInteger,CSBSparseEvent,L1Binary,VerLDPC,SuperImposed,MPCS,BCS}. More specifically, in binary CS, the nonzero elements of the sparse signal have the same value of 1, and the objective is to minimize the number of real valued measurements linearly generated from the binary sparse signal.

The $\ell_1$-minimization solution for the noiseless binary CS has been proposed in \cite{L1Binary}, where problem in (\ref{Relax1s}) is solved by limiting the reconstructed signal in the range [0,1]. Also, by using an integer programming, \cite{UniqueInteger} proposed a model to solve a system of linear equations with integer variables. In \cite{BCS} a simple verification-based decoder has been proposed for binary CS problems, where the measurement matrix is sparse and its elements are randomly obtained based on a predefined probability distribution function. However, the number of required measurements in these approaches are still very large, as they do not take the binary structure of the signal into the design of the CS approaches.

In this paper, we propose a binary CS approach based on analog fountain codes (AFCs) \cite{MahyarLetter}. AFCs were originally designed to approach the capacity of the Gaussian channel across a wide range of signal to noise ratios (SNRs). Each AFC coded symbol is a linear combination of randomly selected information symbols with real weight coefficients. With the proper design of the weight coefficients, each binary linear equation associated with each AFC coded symbol will have a unique solution. As shown in \cite{MahyarLetter}, the AFC decoding problem is equivalent to solving a system of binary linear equations at high SNRs, which is the same as the reconstruction of a binary sparse signal from noiseless linear measurements. This interesting property of AFC codes motivates us to incorporate them in the design of an efficient binary CS approach to minimize the number of measurements required for the successful recovery of the sparse signal.

In the proposed analog fountain compressive sensing (AFCS) approach, each measurement is generated from a number of randomly selected binary information symbols with weight coefficients selected from a finite weight set. Thanks to the random structure of AFC codes, the number of generated measurements is potentially limitless and the transmitter can keep transmitting the measurements till the destination fully recovers the sparse binary signal. As a result, the transmitter in the proposed binary CS can adapt to the time-varying channel condition and thus can maximize the sensing and transmission efficiency. Due to the sparsity and the fact that weight coefficients do not introduce redundancy, each measurement is connected to a very small number of non-zero signal components in the respective weighted bipartite graph; thus, information symbols can be recovered by comparing the measurement value with the weight coefficients assigned to it in an iterative manner. More specifically, we propose a reconstruction algorithm, where each measurement can determine its connected variable nodes if it has connection with at most $T$ nonzero signal elements in the bipartite graph, where $T\ge 0$ is a predefined value. To minimize the number of required measurements, we propose an optimization problem to find the optimum measurement degree, defined as the number of information symbols connected to each measurement or simply the row weight of the measurement matrix. The minimum number of required measurements is then approximated and shown to be of $\mathcal{O}(-n\log(1-s))$, where $n$ is the signal length and $s\triangleq k/n$ is the signal sparsity order. Simulation results show that the proposed scheme has a significantly lower error rate compared to the $\ell_1$-minimization approach \cite{L1Binary} and the binary CS scheme in \cite{BCS}.

We further extend the proposed AFCS scheme to the noisy CS problem, where the measurements are received in the presence of additive white Gaussian noise (AWGN). We propose a message passing decoder to reconstruct the binary sparse signal and show that the AFCS can recover the sparse signal with a very small number of measurements compared to the $\ell_1$-minimization approach \cite{L1Binary}. More specifically, the proposed AFCS scheme requires less number of measurements to completely recover the binary sparse signal in a wide range of SNRs even with a lower measurement degree compared to the $\ell_1$-minimization approach \cite{L1Binary}.

As a direct application of the AFCS, we then consider sparse event detection (SED) in WSNs, where a set of monitoring sensors try to capture a small number of active events randomly distributed in the sensing field \cite{Mahyar_GC_WSN}. CS mechanisms are beneficial to WSNs as sensor nodes usually have constrained sensing, limited memory and processing capability, as well as small communication bandwidth \cite{Weightedl1SED}. Several approaches have used CS theory for SED problem, but they are impractical in some scenarios. For example, Bayesian compressive sensing (BCS) has been used in \cite{BCSnew} to solve the SED problem in WSNs. However, BCS is known to be sensitive to noise, leading to a considerable degraded performance in low SNRs. To solve this problem, \cite{ADLOwWSNS} applied sequential compressive sensing (SCS) to model the SED problem as a support recovery problem and solve it in an adaptive manner. However, in this approach, a large number of sensor nodes (about half that of event sources) is required to be deployed in the sensing field in order to guarantee its performance. This will also bring the system a significant communication cost due to a large number of sensor nodes.

In this paper, we revisit the sparse event detection problem in WSNs from the coding theory perspective and solve the problem as a decoding problem of AFCS. We formulate the variable and check node degree distributions for two scenarios of random and uniform deployment of sensor nodes. The minimum number of sensor nodes required to fully detect all active events in the noiseless case is then found as a function of the sensing coverage radius and the total area of the sensing field. We show that the proposed AFCS scheme approaches this lower bound in high SNRs. We further analyze the probability of false detection for the proposed scheme and show that the AFCS achieves a very low probability of false detection at different SNRs. Simulation results show that the proposed scheme outperforms the SCS \cite{ADLOwWSNS} and BCS \cite{BCSnew} approaches in terms of the probability of correct detection at different SNRs and the minimum sampling ratio to successfully detect the active event sources at different sampling ratios.


Throughout the paper the following notations are used. We use a boldface small letter to denote a vector and the $i^{th}$ entry of vector \textbf{v} is denoted by $v_i$. A boldface capital letter is used to denote a matrix and the $i^{th}$ element in the $j^{th}$ row of matrix \textbf{G} is denoted by $g_{i,j}$. The transpose of matrix \textbf{G} is denoted by $\textbf{G}'$. A set is denoted by a curly capital letter and the cardinality of set $\mathcal{S}$ is represented by $|\mathcal{S}|$.

The rest of the paper is organized as follows. In Section II, we introduce the AFCS approach and the general reconstruction algorithm for the AFCS scheme. The analysis of the proposed reconstruction algorithm in the noiseless case is then presented in Section III, followed by the optimization of the measurement degree and an approximation for the number of required measurements. A message passing decoder is then proposed for AFCS in the presence of additive white Gaussian noise in Section IV. In Section V, we demonstrate the application of AFCS in WSNs for sparse events detection. Section VII provides some discussion on the practical issues for sparse event detection in WSNs. Simulation results are shown in Section VI.  Finally Section VII concludes the paper.

\begin{table}[t]
\caption{Notation Summary}
\label{NotSum}
\centering
\scriptsize
\begin{tabular}{|p{0.75cm}|p{7cm}|}
\hline
\textbf{Notation}&\textbf{Description}\\
\hline
$n$&Length of the binary sparse signal (number of event sources)\\
\hline
$m$&Number of measurements (sensor nodes)\\
\hline
$k$&Number of nonzero elements of the binary sparse signal (number of active events)\\
\hline
$s$& Sparsity order, $s=k/n$\\
\hline
$\beta$& Sampling ratio, $\beta=m/n$\\
\hline
$L$&Measurement (check node) degree\\
\hline
$d_v$&Variable node degree\\
\hline
$\textbf{b}$& binary sparse signal of dimension $n$\\
\hline
$\textbf{G}_{m\times n}$& $m\times n$ measurement matrix, $\textbf{G}=[g_{i,j}]$\\
\hline
$\mathcal{W}$&Weight set of AFCS with size $D$\\
\hline
$T$& Maximum number of weight coefficients included in each operation of the decoding apgorithms\\
\hline
$\mathcal{S}$&Monitoring region of area $S$\\
\hline
$\textbf{e}$& binary sparse event signal of dimension $n$\\
\hline
$\textbf{H}_{m\times n}$&$m\times n$ channel matrix, $\textbf{H}=[h_{i,j}]$\\
\hline
$h_{i,j}$&The channel between the $i^{th}$ sensor node and the $j^{th}$ event source\\
\hline
$d_{i,j}$&The distance between the $i^{th}$ sensor node and the $j^{th}$ event source\\
\hline
$\alpha$& Propagation loss factor\\
\hline
$\epsilon_i$& Thermal measurement noise at the $i^{th}$ sensor\\
\hline
$\lambda$&Received SNR at the sink node\\
\hline
$R_s$& Sensor coverage radius\\
\hline
PCD&Probability of correct detection\\
\hline
PFD&Probability of false detecttion\\
\hline
\end{tabular}
\end{table}
\normalsize
\section{Analog Fountain Compressive Sensing}
We assume that a binary signal vector of length $n$,  \textbf{b}, is strictly $k$-sparse, which means that the number of non-zero elements of \textbf{b} is exactly $k$. The sparsity order of \textbf{b} is then defined as $s\triangleq k/n$. Table \ref{NotSum} provides a summary of notations for quick reference.

Let us first briefly introduce the analog fountain code (AFC).  AFCs were originally proposed in \cite{MahyarWCNC,MahyarLetter} as an effective adaptive transmission scheme to approach the capacity of wireless channels. In AFC,  to generate each coded symbol  $L$ information symbols are selected uniformly at random and then linearly combined with real weight coefficients. For simplicity, we assume that weight coefficients are chosen from a finite weight set, $\mathcal{W}$, with $D\ge L$ positive real members.

Let \textbf{G} denote the generator matrix of dimension $m\times n$, where $m$ is the number of coded symbols, then AFC coded symbols, \textbf{c}, are generated as follows:
\begin{align}
\label{orgAFC}
\textbf{c}=\textbf{Gb},
\end{align}
where, only $L$ elements in each row of matrix \textbf{G} are nonzero and each nonzero element of \textbf{G} is randomly chosen from the weight set $\mathcal{W}$. As shown in \cite{MahyarLetter}, to improve AFC code performance, the generator matrix should have uniform column degrees, which can be simply achieved by selecting information symbols from those which have currently smallest degrees when generating each coded symbol. This way, each information symbol is connected to a fixed number of coded symbols; leading to a regular matrix \textbf{G}.
\subsection{Measurement Matrix Construction}
\begin{figure}[t]
\centering
\includegraphics[scale=0.4]{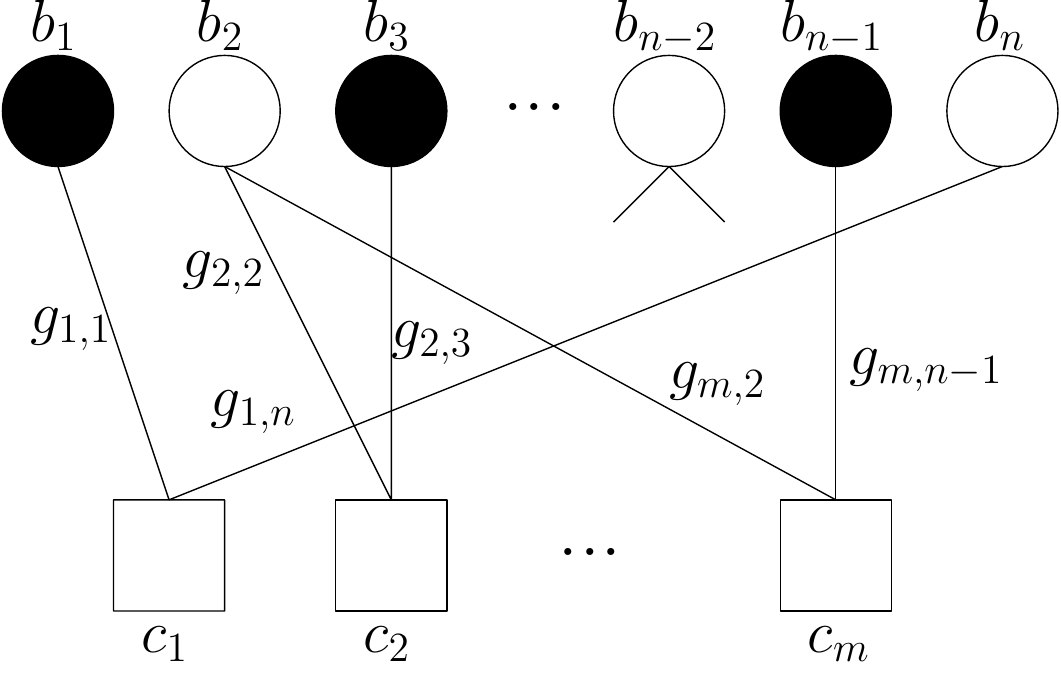}
\caption{The weighted bipartite graph of the AFCS scheme. Black and white circles respectively show the nonzero and zero signal elements.}
\label{bipartiteFig}
\end{figure}

Each row of the measurement matrix is generated in a similar way that a coded symbol is generated in AFCs, that is first an integer $L$, called \emph{measurement degree}, is obtained based on a predefined probability mass function, called \emph{degree distribution}. For simplicity, we assume that the measurement degree, i.e., the number of non-zero elements in each row of the measurement matrix, is a fixed value. Let $\textbf{g}_i$ denote the $i^{th}$ row of the measurement matrix \textbf{G}; then, only $L$ entries of $\textbf{g}_i$ have nonzero values. The value of nonzero elements of $\textbf{g}_i$ are also randomly obtained from the weight set $\mathcal{W}$ without the repetition. Although the weight coefficients can be obtained based on a predefined probability distribution function, for simplicity we assume that the weight set $\mathcal{W}$ has $D\ge L$ positive real members as follows:
\begin{align}
\nonumber\mathcal{W}=\{w_1,w_2,...,w_D\}.
\end{align}
In the next subsection, we show how the weight set $\mathcal{W}$ should be defined. The measurements are then calculated by $\textbf{c}=\textbf{Gb}$, which can be graphically represented by a weighted bipartite graph with elements of the binary signal \textbf{b} and the measurements as variable and check nodes, respectively. In the bipartite graph, the variable node $j$ is connected to the check node $i$, if $g_{i,j}$ is non-zero, and the edge between them is also assigned with the value of $g_{i,j}$. Fig. \ref{bipartiteFig} shows the weighted bipartite graph of the proposed binary CS scheme, where the measurement degree is $L=2$.

As we consider that the locations of non-zero elements in each row of the measurement matrix are selected quite randomly, it is highly probable that a variable node is not connected to any check node in the corresponding bipartite graph, especially when the number of measurements is relatively low.  More specifically, the probability that a variable node is not connected to any check node is $e^{-mL/n}$ \cite{RegLT}. This problem significantly increases the error rate for the CS schemes in \cite{BCS,L1Binary}.

To overcome this problem, we modify the measurement matrix in a way that each variable node is connected to at least one check node. Moreover, to ensure that all the variable nodes have the same chance to be recovered, the degrees of all variable nodes are chosen to be the same. To achieve this, we select the variable nodes among those with the lowest degrees when generating each measurement. In this way, the variable node degrees will be either $d_v$ or $d_v-1$, where $d_v$ is the smallest integer larger than or equal to $mL/n$ \cite{RegLT}. Later in Section V, we show how this imposed structure in the measurement matrix can be obtained in real applications, which significantly increases the sensing efficiency and improve the reconstruction capability. It is important to note that the measurement matrix is not fixed and randomly generated for each experiment. In fact, the measurement matrix is generated in a way to have uniform column and row degrees, where the nonzero elements of the measurement matrix are selected from the predefined weight set.
\subsection{Weight Set Design}
It is important to note that the weight set should be designed such that each measurement can uniquely determine its connected variable nodes. For instance, let us assume that $w_1+w_2=w_3$ and a measurement has a value of $w_3$ and is connected to $b_1$, $b_2$, and $b_3$ with weight coefficients, $w_1$, $w_2$, and $w_3$, respectively. The decoder then try to solve the binary linear equation $w_1b_1+w_2b_2+w_3b_3=w_3$. This measurement can determine either $b_1=b_2=1, b_3=0$, or $b_1=b_2=0, b_3=1$, i.e., two possible solutions. To guarantee that each measurement can uniquely verify its connected variable nodes, it is sufficient that the weight set $\mathcal{W}$ satisfies the following condition:
\begin{align}
\label{WeightCond}
\sum_{i=1}^{D}v_iw_i\ne 0,
\end{align}
for all $\textbf{v}\in\{-1,0,1\}^D$, where $\textbf{v}\ne\{0\}^D$, and $w_i\in\mathcal{W}$ for $i=1,...,D$. As stated in \cite{MahyarLetter}, each measurement is associated with a binary linear equation and thus the decoder task is actually to solve a system of binary linear equations in the noiseless case. Condition (\ref{WeightCond}) guarantees that the linear equation associated with each measurement has a unique solution.

Here, we assume that the nonzero elements of the measurement matrix are sampled from a finite weight set, where the weight set satisfies (\ref{WeightCond}). It is clear that when the members of the weight set are sampled from a continuous probability distribution function, (\ref{WeightCond}) can be easily satisfied. In practice, the elements of the measurement matrix depend on the physical condition of the system, which is mainly random. For example, as we will later see in Section V, in the sparse event detection in WSNs, the elements of the measurement matrix are channel gains, which are random due to shadowing and small scale fading; thus, condition (\ref{WeightCond}) is satisfied and each measurement can uniquely determine its connected variable nodes. In the rest of the paper, we use weight coefficients which have been obtained from a zero mean Gaussian distribution with variance 1, unless otherwise specified.

It is important to note that in conventional CS theory,  restricted isometry property (RIP) is sufficient to guarantee sparse reconstruction by $\ell_1$-minimization approach \cite{CSDonoho}. It is also known that $\ell_1$-minimization reconstructs every sparse signal precisely when the sensing matrix satisfies the null space property (NSP) \cite{cohen2009compressed}. In general, the null space of measurement matrix \textbf{G} is defined as $\mathcal{N}(\textbf{G})=\{\textbf{z}|\textbf{Gz}=\textbf{0}\}$. If we want to recover all sparse signals \textbf{x} from the measurements \textbf{Gx}, it is clear that for any pair of distinct vectors \textbf{x} and $\textbf{x}'$, we must have $\textbf{Gx}\ne\textbf{Gx}'$, since otherwise it would be impossible to distinguish \textbf{x} from $\textbf{x}'$. In other words, to recover all sparse signals from the measurements generated by matrix \textbf{G}, we must have $\mathcal{N}(\textbf{G})=\{\textbf{0}\}$, where \textbf{0} is the zero vector of dimension $n$. It is worth mentioning that if the weight coefficients in AFCS satisfy condition (\ref{WeightCond}), then each sparse binary signal is uniquely recovered form its linear measurements. In other words,  the binary structure of the sparse signal enables us to design the measurement matrix which can precisely satisfy the null sparse property ($\mathcal{N}(\textbf{G})=\{\textbf{0}\}$) for all binary signals, which is sufficient for compressive sensing.
\subsection{Verification Based Reconstruction}

It is clear from (\ref{WeightCond}) that a zero valued measurement can uniquely verify its connected variable nodes as zero signal elements. In fact, condition (\ref{WeightCond}) guarantees that a binary linear equation with weight coefficients from $\mathcal{W}$ has a unique solution of zero if and only if the binary variables are all zero \cite{VerifyCS,Verify}. Also due to signal sparsity, only few nonzero signal elements are connected to each measurement. Therefore, by comparing the value of a measurement and the sum of different numbers of its weight coefficients, we can uniquely determine the values of its connected variable nodes. Based on these observations, we then propose the following verification based reconstruction algorithm, referred to as the \textit{Sum Verification Decoder}.

Let us consider the binary compressive sensing problem $\textbf{c}=\textbf{Gb}$, where the binary sparse signal vector \textbf{b}$_{n\times 1}$ has to be reconstructed from linear measurements \textbf{c}$_{m\times 1}$ and a given measurement matrix \textbf{G}$_{m\times n}$. We use the notation $\mathcal{N}_i$ to denote the set of variable nodes connected to check node $c_i$, and use $\mathcal{N}_i\backslash  \mathcal{B}$ to denote the set of connected variable nodes to check node $c_i$, excluding the variable nodes in the set $\mathcal{B}$. The operations of the \textit{Sum Verification Decoder} can then be summarized as follows, for a given $T\ge 0$:

\textbf{Sum Verification Decoder}: Construct the weighted bipartite graph from the measurement matrix \textbf{G}. Construct a matrix $\textbf{G}_T$, where its rows correspond to all binary vectors of dimension $D$ with at most $T$ nonzero elements. Construct a vector $\textbf{w}=[w_1,w_2,...,w_D]'$. Compute the vector \textbf{u} as follows: $\textbf{u}=\textbf{G}_T\textbf{w}$. Initially, we assume that \textbf{b} is a zero vector of dimension $n$.
\begin{itemize}
\item[1.] Find a measurement $c_i$ with a nonzero degree which has a value equal to one of the elements of vector \textbf{u} (if there is no such measurement, this decoding algorithm halts at this point and fails to recover all the variable nodes). Find the set of its connected variable nodes $\mathcal{N}_i$.
    \begin{itemize}
\item[a)] Find the number of elements in $\mathcal{N}_i$, i.e. $|\mathcal{N}_i|=\tau$, where $\mathcal{N}_i=\{j_1,...,j_{\tau}\}$. Construct a matrix \textbf{$\Gamma$}, where its rows correspond to all binary vectors of dimension $\tau$ with at most $T$ nonzero elements. Construct a vector $\textbf{g}^{(1)}_i=[g_{i,j_1},g_{i,j_2},...,g_{i,j_{\tau}}]'$. Compute the vector \textbf{z} as follows: $\textbf{z}=$\textbf{$\Gamma$}$\textbf{g}^{(1)}_i$. Find $\ell$ such that $z_{\ell}=c_i$.
\item[b)] Set $b_{j_r}=\Gamma_{\ell,r}$ for $r=1,2,...,\tau$.
\end{itemize}
\item[2.] Remove all the edges connected to verified variable nodes $b_{j_r}$.
\item[3.] Update the measurement values to $\textbf{c}=\textbf{c}-\textbf{Gb}$.
\item[4.] Repeat (1), (2), and (3) until all $b_j$s are determined
\end{itemize}

In each iteration of the proposed decoder, a measurement which has a connection to at most $T$ nonzero variable nodes is found (Step 1). Since the weights are chosen from a fixed weight set, this can be achieved by simply comparing the value of the measurements with a table consisting of all the possible values of the summation of at most $T$ weight coefficients. If such a measurement cannot be found, the decoder will be terminated and the variable nodes cannot be fully recovered. In the next step, the variable nodes that are connected to the selected measurement will be uniquely recovered (Step 1, (a) and (b)). Then, the edges connected to the recovered variable nodes are removed from the bipartite graph (Step 2). Also, the measurement values will be updated to cancel out the effect of the recovered variable nodes on the remaining measurements (Step 3). This algorithm is repeated until all the variable nodes are recovered.

\begin{figure}[t]
\centering
\includegraphics[scale=0.45]{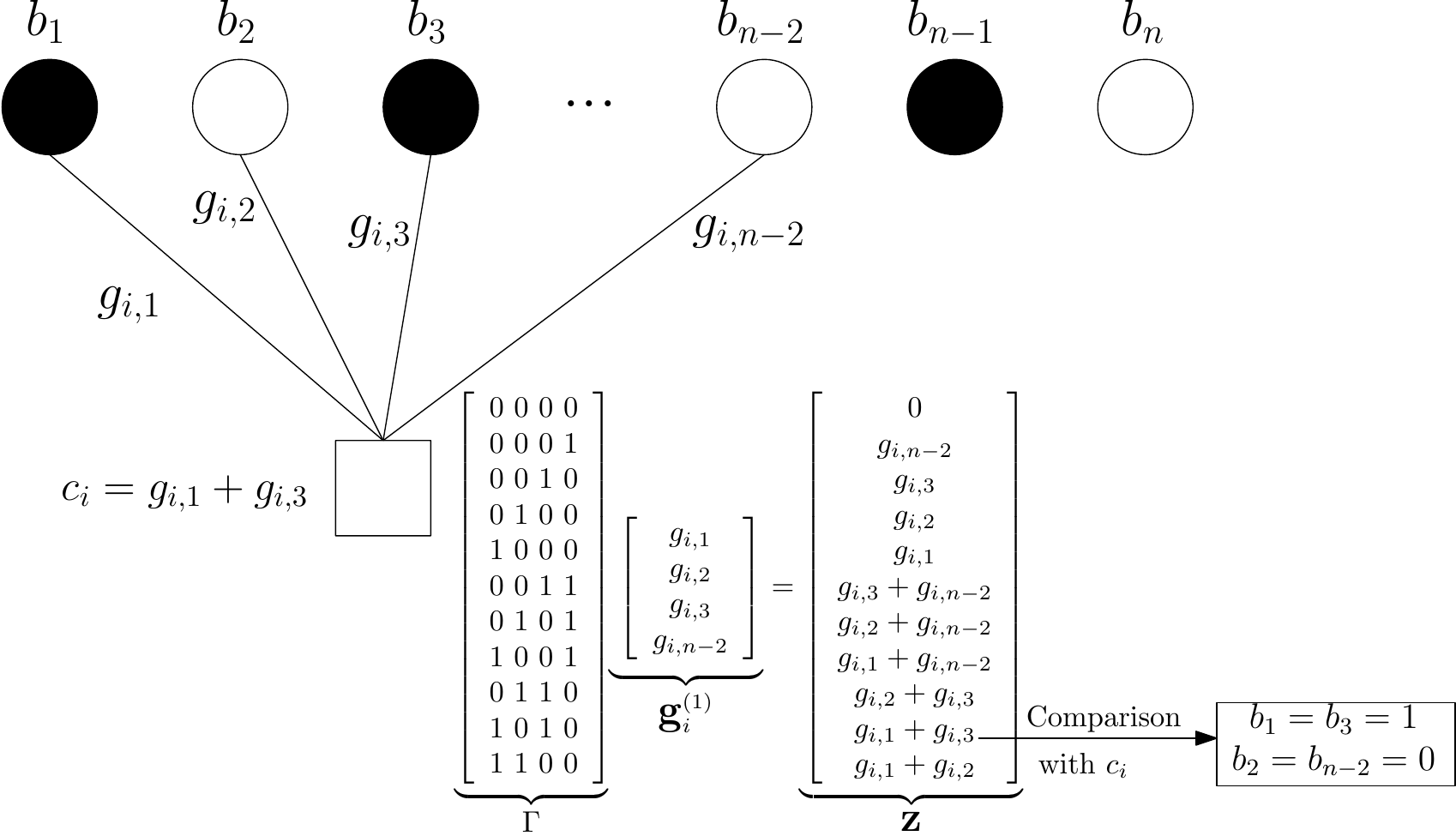}
\caption{The sum-verification reconstruction algorithm for $L=4$ and $t=2$. Black and white circles show the non-zero and zero signal elements, respectively.}
\label{fig2deg}
\end{figure}

Since the signal is sparse, the probability that a measurement is connected to a large number of nonzero elements of the signal is very low. In other words, a majority of  measurements are connected to a very small number of nonzero variable nodes. Therefore, it is reasonable to only consider small values of $T$, which also avoids huge complexity at the decoder. More specifically, in this paper we only consider $T=0$, $1$, and $2$. Fig. \ref{fig2deg} shows the proposed reconstruction algorithm for the case that the measurement degree is $L=4$ and $T=2$. As can be seen in this figure, measurement $c_i=g_{i,1}+g_{i,3}$ is connected to $b_1$, $b_2$, $b_3$, and $b_{n-2}$, and by comparing the measurement value with vector $\textbf{z}$, all connected variable nodes can be uniquely determined.
\section{Performance Analysis of the Verification Based Decoder}
In this section, the performance of the sum verification decoder is analyzed for different values of $T$. We then find the optimum measurement degree as a function of the sparsity order, $s$. An upper bound is also derived for the number of measurements required to successfully recover all variable nodes.
\subsection{Analysis of the Verification Probability}

\begin{figure}[t]
\centering
\includegraphics[scale=0.33]{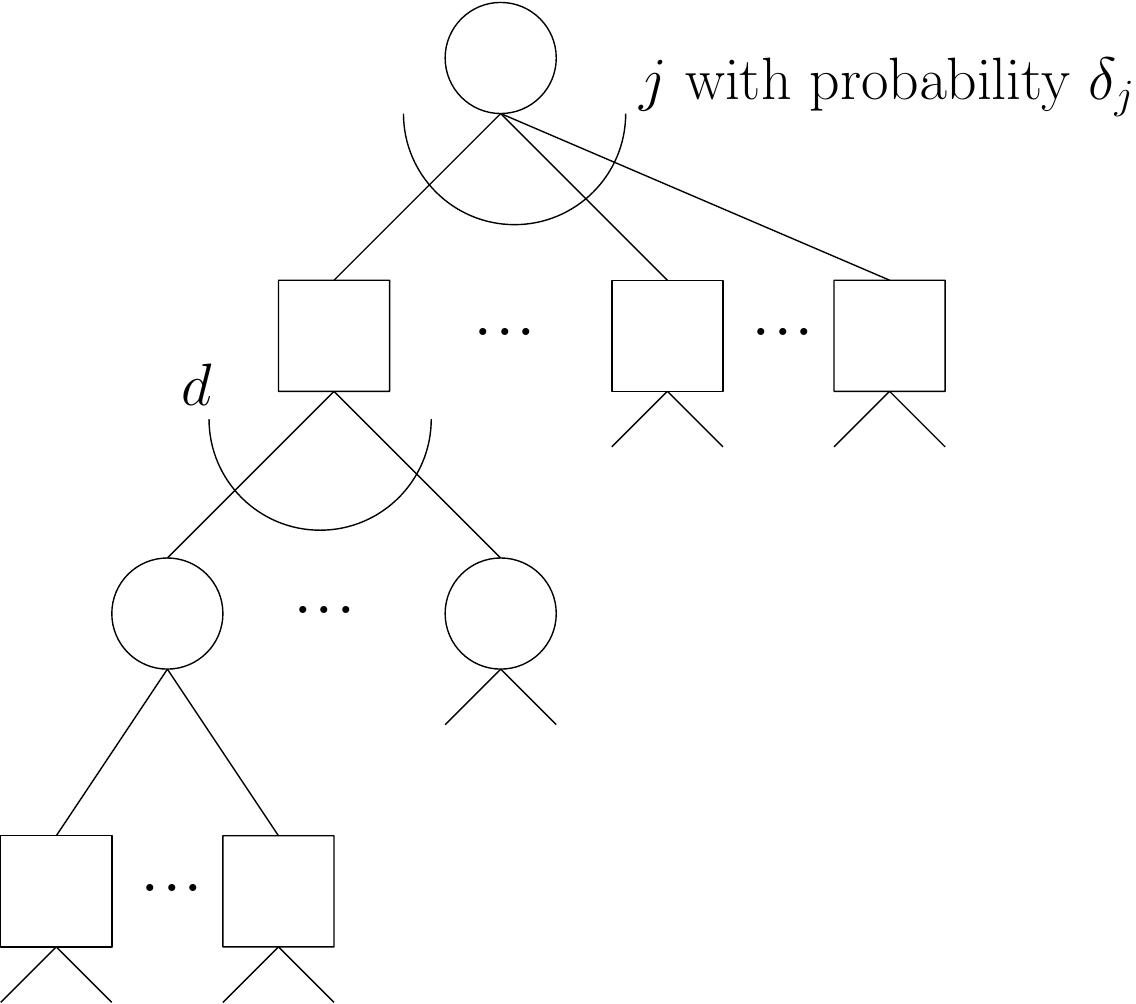}
\caption{SUM-OR tree $G_{\ell}$ when $d$ and $j$ indicate the number of edges connected to the specific node. Circles and squares show OR and SUM nodes, respectively.}
\label{SUMORfig}
\end{figure}

Let $\mathcal{G}$ denote the bipartite graph corresponding to the code at the receiver. We choose subgraph $\mathcal{G}_\ell$ as following. Choose an edge $(v,w)$ uniformly at random from all edges. Call the variable node $v$ the root of $\mathcal{G}_\ell$. Subgraph $\mathcal{G}_\ell$ is the graph induced by $v$ and all neighbors of $v$ within distance $2\ell$ after removing the edge $(v;w)$. As shown in \cite{Raptor}, $\mathcal{G}_\ell$ is a tree asymptotically, so we refer to $\mathcal{G}_l$ as a SUM-OR tree,  where SUM and OR nodes refer to check and variable nodes, respectively. Fig. \ref{SUMORfig} shows the SUM-OR tree of depth $2\ell $, where the root of the tree is at depth $0$ and has children at depth $1$. More specifically, a node at depth $t$ has children at depth $t+1$ for $t=0,...,2\ell-1$ and nodes at depths $0,2,..,2\ell$ are referred to as OR nodes, while nodes at depths $1,3,...,2\ell-1$ are referred to as SUM nodes.

In the following, we rephrase the proposed sum verification decoder by using the SUM-OR tree. At every iteration of the algorithm, messages (0 or 1) are passed along the edges from check nodes to variable nodes, and then from variable nodes to check nodes. A variable node sends 0 to an adjacent check node if and only if its value is not recovered yet. Similarly, a check node sends 0 to an adjacent variable node if and only if it is not able to recover the value of the variable node. In other words, a variable node sends 1 to a neighboring check node if and only if it has received at least one message with value 1 from its other neighboring check nodes. Also, a check node sends 0 to a neighboring variable node if and only if it is connected to more than $T$ unrecovered variable nodes.

Since, a subgraph expanded from each variable node is asymptotically a tree, then the message exchange between variable and check nodes in the bipartite graph can be seen as the message exchange between SUM and OR nodes in the tree. More specifically, the messages from OR nodes at depth $2i$ are passed to SUM nodes at depth $2i-1$, which calculate new messages to forward to OR nodes at depth $2i-2$, corresponding to one iteration of the decoding algorithm. Therefore, if we map the first iteration of the decoding algorithm to the message exchange between OR and SUM  nodes at depths $2\ell$ and $2\ell-1$, respectively, then after $\ell$ iterations, the root of the tree will receive messages from its children. The decoding error probability is then the probability that the root of the tree does not receive at least one message of value 1 from its children. The following lemma gives the decoding probability which is equivalent to the probability that the root of the SUM-OR tree does not receive a message of value 1 from its children.

\newtheorem{lemma}{Lemma}
\begin{lemma}
\label{sumortreelemma}
We consider the binary AFCS, where the measurements are generated by the AFCS process. Let $p_i$ denote the probability that a variable node is recovered after $i$ iterations of the sum verification decoder, where $p_0=0$. Let $q_{i}^{(0)}$ and $q_{i}^{(1)}$ denote the probability that an unrecovered variable node in the $i^{th}$ iteration of the reconstruction algorithm is zero and one, respectively, where $q_0^{(i)}=1-q_1^{(i)}$. Then for a fixed measurement degree $L$, $p_i$ is calculated as follows:
\begin{align}
\label{finalver}
p_i=1-\delta(1-f_i),
\end{align}
where $\delta(x)=\sum_{j}\delta_jx^j$,
\begin{align}
\nonumber f_i&=\sum_{i=0}^{d}\sum_{j=0}^{A}\dbinom{d}{i}\dbinom{d-i}{j}p_{i-1}^i\left(q_{i-1}^{(0)}(1-p_{i-1})\right)^{d-i}\lambda_{i-1}^j\\
&+\sum_{i=0}^{d-T}\dbinom{d}{i}\dbinom{d-i}{T}q_{i-1}^{(0)}p_{i-1}^i\left(q_{i-1}^{(0)}(1-p_{i-1})\right)^{d-i}\lambda_{i-1}^T,
\label{lemmapc}
\end{align}
$d=L-1$, $\lambda_{i}=q_{i}^{(1)}/q_{i}^{(0)}$, $A=\min\{T-1,d-i\}$, $q_{i}^{(0)}=1-\frac{k}{n}$, and $\delta(x)$ is the variable node degree distribution function with respect to the edges.
\end{lemma}

The proof of this lemma is provided in Appendix \ref{proofsumor}. To complete the analysis, we only need to find $\delta(x)$. As we assume that each measurement is of degree $L$, variable nodes are either of degree  $d_v$ or $d_v-1$, where $d_v$ is the smallest integer larger than or equal to $\beta L$ and $\beta$ is the sampling ratio, defined as the ratio of the number of measurements and that of variable nodes. Let $v_1$ and $v_2$ denote the probability that a variable node is of degree $d_v$ and $d_v-1$, respectively, then $v_1+v_2=1$ and
\begin{align}
v_1d_v+v_2(d_v-1)=\beta L,
\end{align}
which results in $v_1=1-d_v+\beta L$ and $v_2=d_v-\beta L$. Thus, the variable node degree distribution, $\Delta(x)$, can be written as follows:
\begin{align}
\nonumber\Delta(x)=(1-d_v+\beta L)x^{d_v}+(d_v-\beta L)x^{d_v-1}.
\end{align}
Accordingly, $\delta(x)$, can be calculated as follows \cite{Raptor}:
\begin{align}
\nonumber\delta(x)=\frac{\Delta'(x)}{\Delta'(1)}=\frac{v_1d_vx^{d_v-1}+v_2(d_v-1)x^{d_v-2}}{\beta L},
\end{align}
where $\Delta'(x)$ is the derivative of $\Delta(x)$ with respect to $x$.

It is important to note that if $T=L$, then according to (\ref{lemmapc}) we have $f_i=1$, and accordingly $p_i=1$. This means that each check node can uniquely determine its connected variable nodes if the parameter $T$ equals to the measurement degree in the sum verification decoder.
\subsection{Optimization of the Measurement Degree}
According to Lemma \ref{sumortreelemma}, we can formulate an optimization problem to find the optimum measurement degree in order to minimize the number of measurements required for the successful recovery of all variable nodes. However, as can be seen in this lemma, the verification probability is not a linear function of the measurement degree; thus, finding the optimum degree is not straightforward. To overcome this problem, we propose a suboptimum solution to find the measurement degree to maximize the number of recovered variable nodes in each iteration of the sum verification decoder. More specifically, in the first iteration of the decoding algorithm, we have $p_0=0$, $q^{(0)}_0=1-k/n$, and $\lambda_0=\frac{k}{n}/(1-\frac{k}{n})$; thus, we have:
\begin{align}
\label{firstmeas}
f_1=\dbinom{d}{T}\left(q_{0}^{(0)}\right)^{L}\lambda_{0}^T+\sum_{j=0}^{T-1}\dbinom{L-1}{j}\left(q_{0}^{(0)}\right)^{L-1}\lambda_{0}^j.
\end{align}
In other words, a measurement can determine its connected variable nodes if the number of nonzero variable nodes connected to that measurement is at most $T$. Thus, (\ref{firstmeas}) can be rewritten as follows:
\begin{align}
f_1=\sum_{j=0}^{T}\dbinom{L}{j}\left(q_{0}^{(0)}\right)^{L-j}\left(q_{0}^{(1)}\right)^{j}.
\end{align}
Let $q_0\triangleq q_{0}^{(0)}$ which equals to $1-\frac{k}{n}$, then the average number of variable nodes, $R$, which can be verified by each measurement, can be calculated as follows:
\begin{align}
\label{snew}
R=L\sum_{j=0}^{T}\dbinom{L}{j}q_0^{L-j}(1-q_{0})^{j},
\end{align}
The following lemma gives the optimum measurement degree, $L_{opt}$, for a given $T$ and $s$.
 \begin{figure}[t]
\centering
\includegraphics[scale=0.34]{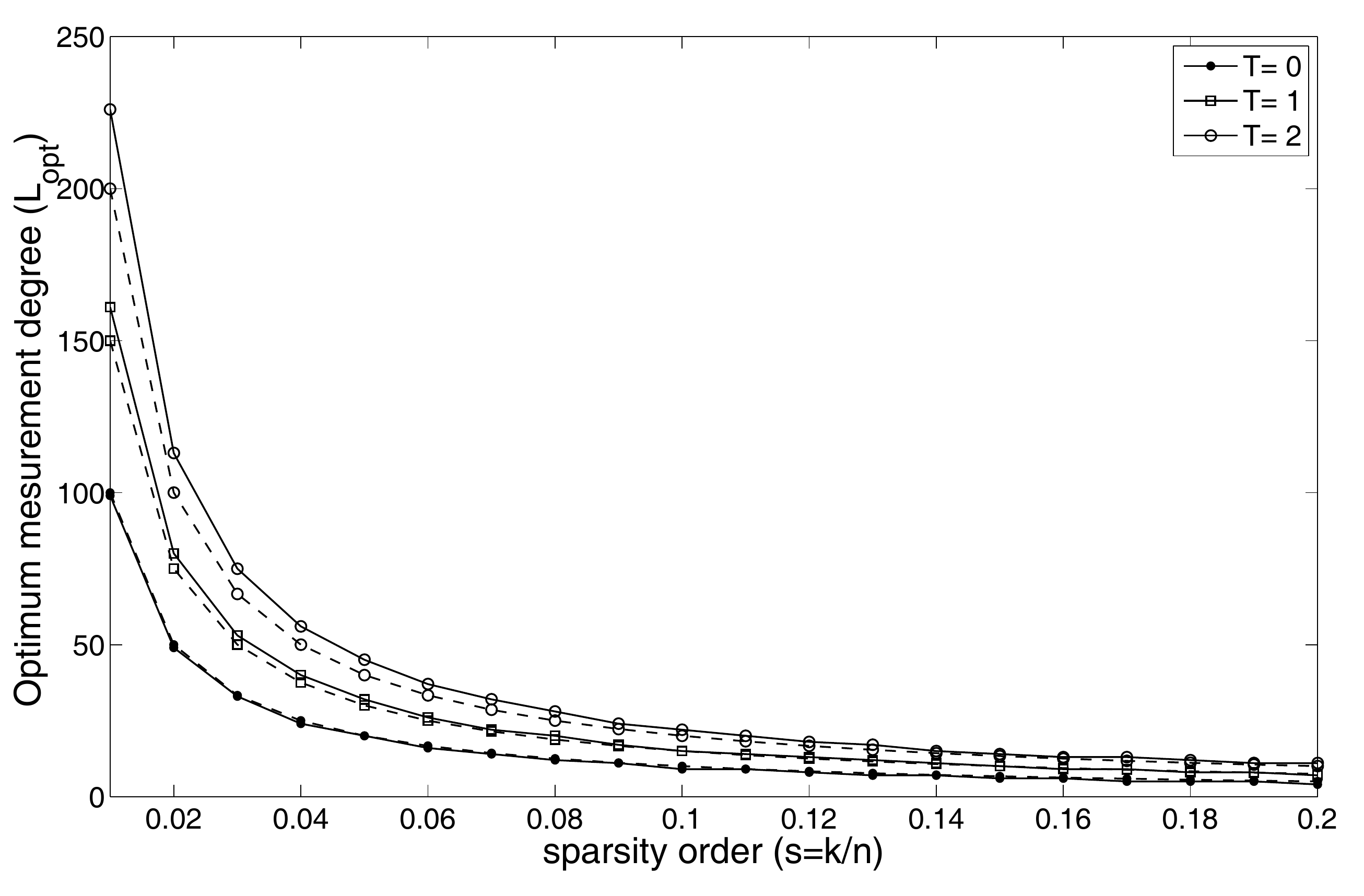}
\caption{The optimum measurement degree versus the sparsity order of the signal. Solid lines show the optimum values, and dashed line shows the approximation which has been obtained from (\ref{Loptlemma}).}
\label{loptfig}
\end{figure}
\begin{lemma}
\label{OptLlemma}
The optimal measurement degree, $L_{opt}$, to maximize the average number of variable nodes, which can be verified by each measurement for a given $T$ and sparsity order $s$ in the sum verification decoder, can be approximated as follows:
\begin{align}
\label{Loptlemma}
L_{opt}\approx\lceil-\frac{T+2}{2\log(1-s)}\rceil,
\end{align}
where $\lceil . \rceil$ is the ceil operator.
\end{lemma}

The proof of this lemma is provided in Appendix \ref{proofOptLemma}. From (\ref{Loptlemma}) it can be noted that the optimum measurement degree is a decreasing function of the sparsity order, $s$, as $|\log(1-s)|$ is an increasing function of $s$. This means that to guarantee that a measurement can verify the maximum number of variable nodes form a very sparse signal, the measurement degree should be very large. Moreover, a binary signal with a higher $s$ require a lower measurement degree, as the number of nonzero variable nodes is relatively high; thus, each measurement is connected to a relatively large number of nonzero variable nodes. Fig. \ref{loptfig} shows the optimized measurement degrees for different sparsity orders and different values of $T$. 

\subsection{The number of measurements}
In the proposed reconstruction algorithm, we can verify a large number of variable nodes, including zero and nonzero signal components, from each measurement because a majority of the measurements are connected to a very small number of non-zero elements of the sparse signal. The following lemma gives an upper bound on the number of measurements required to recover the sparse signal by using the proposed reconstruction algorithm (see Appendix \ref{prooforder} for the proof).
\begin{lemma}
\label{orderlemma}
We consider the binary AFCS, where the measurements are generated by the AFCS process with the measurement degree $L_{opt}$ calculated in (\ref{Loptlemma}). Let $m$ denote the minimum number of measurements required for the signal of length $n$ and sparsity order $s$, to be completely reconstructed by the sum verification decoder. Then, $m$ is of $\mathcal{O}\left(-n\log(1-s)\right)$.
\end{lemma}
It can be easily proven that the propose scheme requires fewer measurements compared to the existing binary CS approach \cite{BCS}, which the number of required measurements is of $\mathcal{O}(k\log (n))$. This is because $\log(x)/x$ is a decreasing function of $x$ when $x\ge e$; thus, $-n \log(1-s)\le k\log(n)$ for $k\le n-e$.
\subsection{Performance Evaluation of noiseless AFCS}
\begin{figure}[t]
\centering
\includegraphics[scale=0.35]{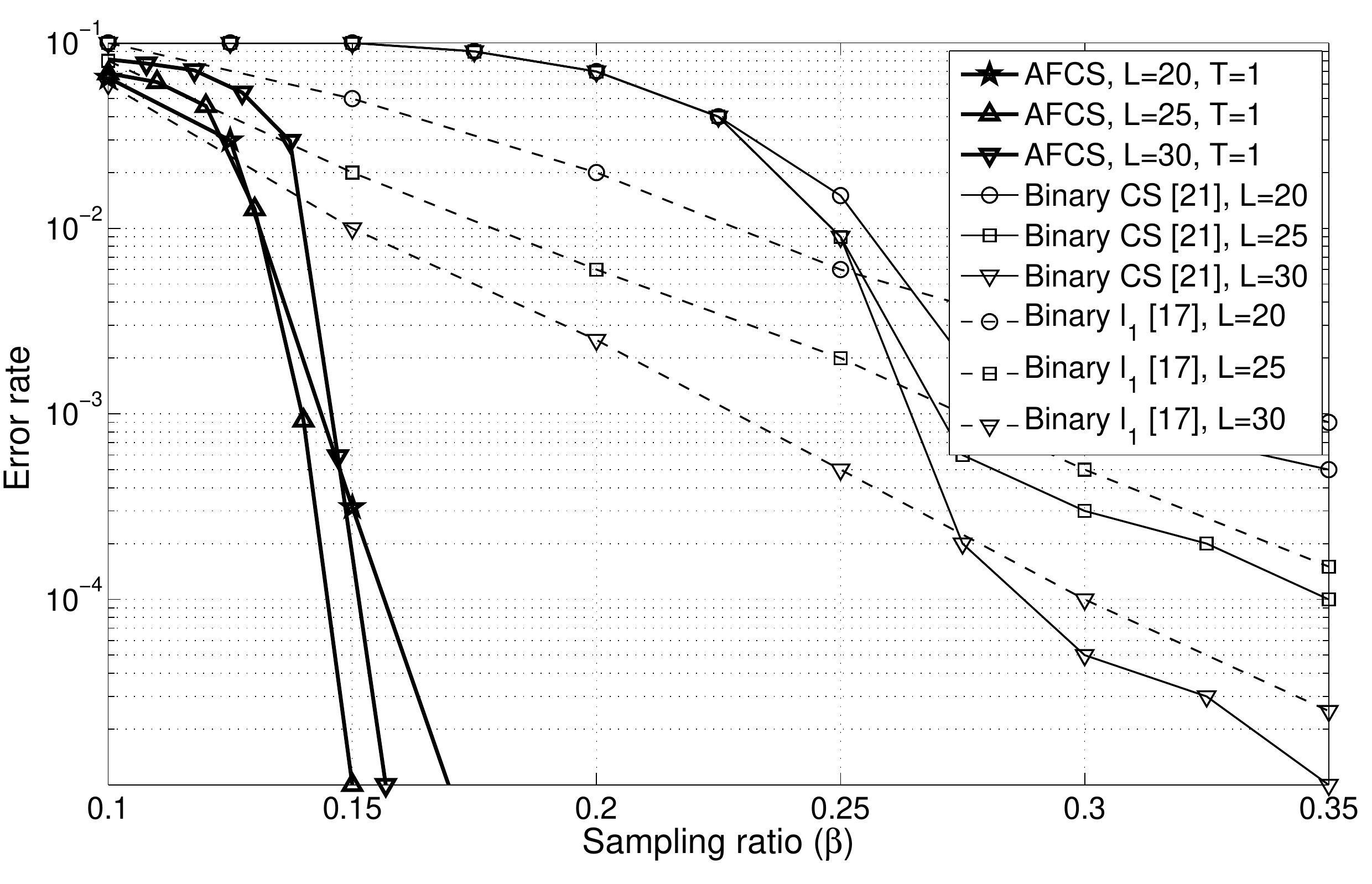}
\caption{The error rate versus the sampling rate for the sparse signal of length $n=1000$, and the sparsity order $s=0.1$.}
\label{afcsvsbcs}
\end{figure}
Fig. \ref{afcsvsbcs} shows the error rate versus the sampling ratio for the proposed AFCS scheme, the $\ell_1$-minimization approach \cite{L1Binary}, and the binary CS approach in \cite{BCS}, when the signal length is $n=1000$ and the sparsity order is $s=0.1$. Here, the error rate is defined as the ratio of the number of unrecovered signal components and the signal length $n$. It is important to note that the decoding process in \cite{BCS} is similar to the sum verification decoding algorithm with $T=1$. As can be seen in this figure, at sampling ratio of $0.15$, the AFCS scheme with $L=20$ achieves the error rate of $10^{-4}$; however, the error rate of the binary CS approach in \cite{BCS} with the measurement degree $L=30$ is in the order of $10^{-1}$, and the error rate in binary $\ell_1$-minimization \cite{L1Binary} is in the order of $10^{-2}$.  In other words, the AFC scheme with an even smaller measurement degree outperforms the existing binary CS schemes. The reason is that in $\ell_1$-minimization approach and binary CS scheme in \cite{BCS}, the variable nodes are selected uniformly at random to generate each measurement; thus, even with a large number of measurements some variable nodes may still not be selected by any measurement, leading to a higher error rate. However, in AFCS, the variable node degrees are uniform so the probability that a variable node is not selected by any check node is zero. Thus, AFCS can approach very small error probabilities even with a smaller measurement degree and decoding parameter $T$. Furthermore, as can be seen in Fig. \ref{afcsvsbcs}, the proposed AFCS approach with $L=25$ has a lower error rate compared to the AFCS scheme with $L=20$ and $L=30$, due to the fact that the optimum measurement degree for sparsity order $s=0.1$ and $T=1$ is about $L=25$; thus, with measurement degrees higher or lower than $L=25$, the error performance will be degraded.
\begin{figure}[t]
\centering
\includegraphics[scale=0.36]{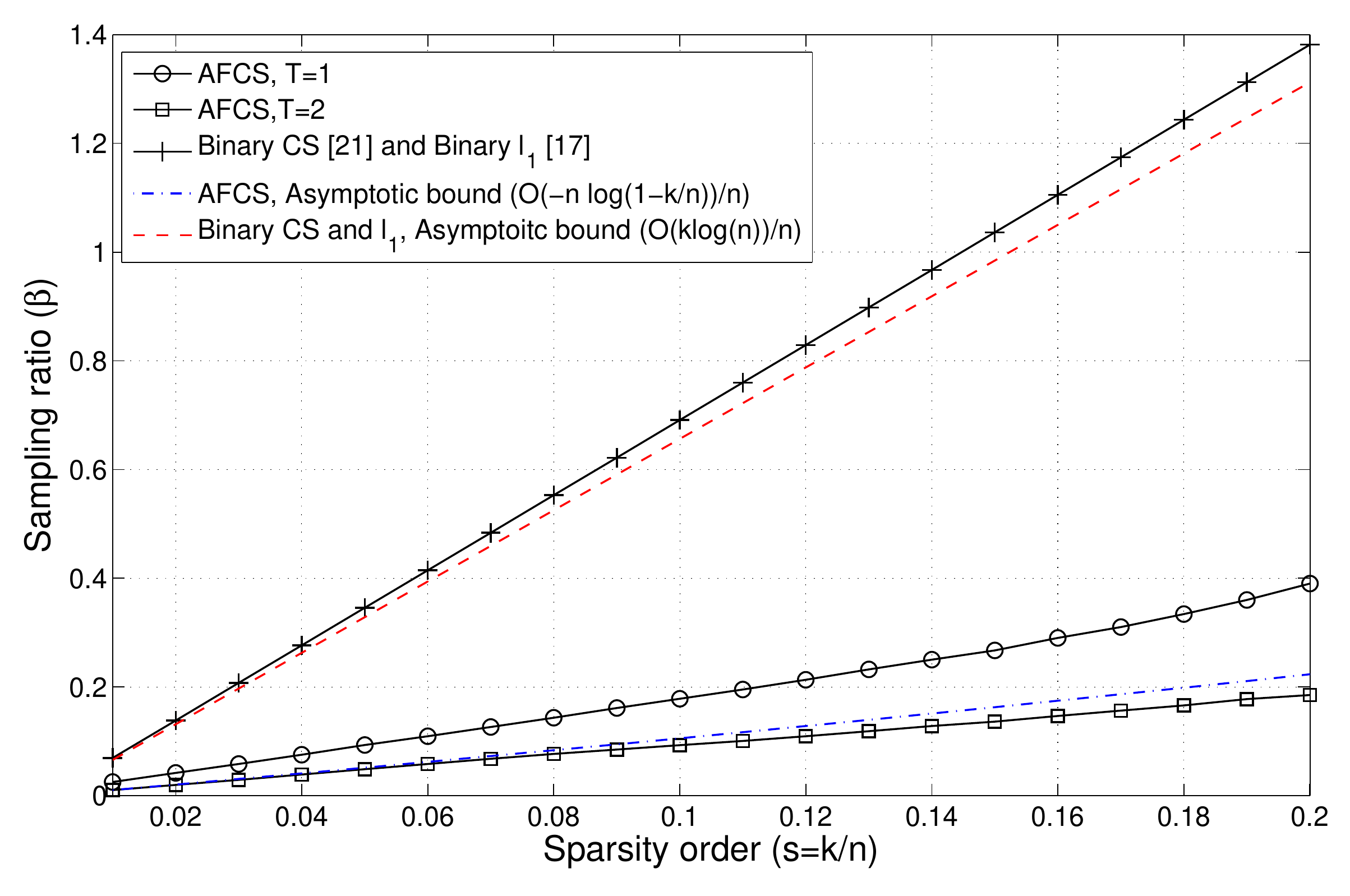}
\caption{The minimum sampling ratio required to successfully reconstruct a binary sparse signal of length $n=1000$ versus the sparsity order.}
\label{simm}
\end{figure}

Fig. \ref{simm} shows the number of measurements required to successfully recover the binary sparse signal versus the sparsity order. As can be seen in Fig. \ref{simm}, the number of measurements for the proposed AFCS scheme is significantly lower than that for the binary CS scheme in \cite{BCS}. In fact, the number of measurements for the binary CS scheme in \cite{BCS} is of $\mathcal{O}(k\log(n))$ and that for the proposed AFCS scheme is of $\mathcal{O}(-n\log(1-\frac{k}{n}))$.

It is important to note that in the proposed reconstruction algorithm, the maximum value of $T$ is considered to be 2. Thus, for each measurement the total number of $1+L+\dbinom{L}{2}$ comparisons should be performed to recover the connected variable nodes. As a result, the total number of comparisons is of $\mathcal{O}(L^2)$. As we have shown that the number of required measurements is of $\mathcal{O}(-n\log(1-s))$ and the optimum measurement degree is $L_{opt}\approx -1/\log(1-s)$, then the reconstruction complexity of the proposed algorithm is of $\mathcal{O}(-n/\log(1-s))$. This indicates that the complexity of the proposed algorithm increases when the number of nonzero elements of the sparse signal is very small. In fact when $s$ goes to zero, $-1/\log(1-s)$ can be approximated by $1/s$, so the complexity of the decoder will be of $\mathcal{O}(n^2/k)$.
\section{Noisy Binary Compressive Sensing}
In this section, we consider the problem of binary compressive sensing in the presence of additive white Gaussian noise (AWGN). In the presence of the noise, AFCS measurements are given by
\begin{align}
\textbf{y}=\textbf{Gb}+\textbf{z},
\end{align}
where \textbf{z} is an $m$ by 1 vector of zero mean additive white Gaussian noise with variance $\sigma^2_z I_m$ and $I_m$ is the $m\times m$ identity matrix.  In \cite{MahyarLetter}, we proposed a belief propagation (BP) decoder for analog fountain codes, which is actually a modification of the decoding algorithm originally proposed for sparse signal reconstruction in \cite{BaronSarvothamBraniuk}. Here, we use the same decoder to reconstruct the binary sparse signal in the presence of additive white Gaussian noise.
In each iteration of the BP decoder, messages (conditional probabilities) are passed from check to variable nodes and then from variable to check nodes. Let $q^{(\ell)}_{ji}(0)$ and $q^{(\ell)}_{ji}(1)$ denote the messages that are passed from variable node $j$ to check node $i$ in the $\ell^{th}$ iteration of the message passing algorithm. Similarly, the messages passed from check node $i$ to variable node $j$ in the $\ell^{th}$ iteration are denoted by $m^{(\ell)}_{ij}(0)$ and $m^{(\ell)}_{ij}(1)$. Message $m^{(\ell)}_{ij}(n)$ is calculated as follows for $r\in\{0,1\}$:
\begin{gather}
\nonumber m^{(l)}_{ij}(r)=p(b_j=r|y_i)=p(\sum_{j'\in\mathcal{N}(i)\backslash j}b_{j'}g_{i,j'}=y_i-rg_{i,j}),
\end{gather}
where  $\mathcal{N}(i)\backslash j$ is the set of all neighbors of check node $i$ except variable node $j$. Also, message $q^{(\ell)}_{ji}(r)$ is calculated as follows.
\begin{gather}
\nonumber q^{(\ell)}_{ji}(r)=C_{ji}\prod_{i'\in\mathcal{M}(j)\backslash i} m^{(\ell)}_{i'j}(r),~r\in\{0,1\}
\end{gather}
where $\mathcal{M}(j)\backslash i$ is the set of all neighbors of variable node $j$ except check node $i$, and $C_{ji}$ is chosen in a way that $q^{(\ell)}_{ji}(0)+q^{(\ell)}_{ji}(1)=1$. After a predefined number of iterations, called $F$, the final probabilities for variable nodes are given by:
 \begin{gather}
\nonumber p^{(F)}_{j}(r)=C_{j}\prod_{i\in\mathcal{M}(j)} m^{(F)}_{ij}(r),
 \end{gather}
 where $C_j$ is chosen in a way that $p^{(F)}_{j}(0)+p^{(F)}_{j}(1)=1$. Variable node $b_j$ is then decided to be $0$ if $p^{(F)}_{j}(0)>p^{(F)}_{j}(1)$; otherwise, it is set to $1$.

\begin{figure}[t]
\centering
\includegraphics[scale=0.31]{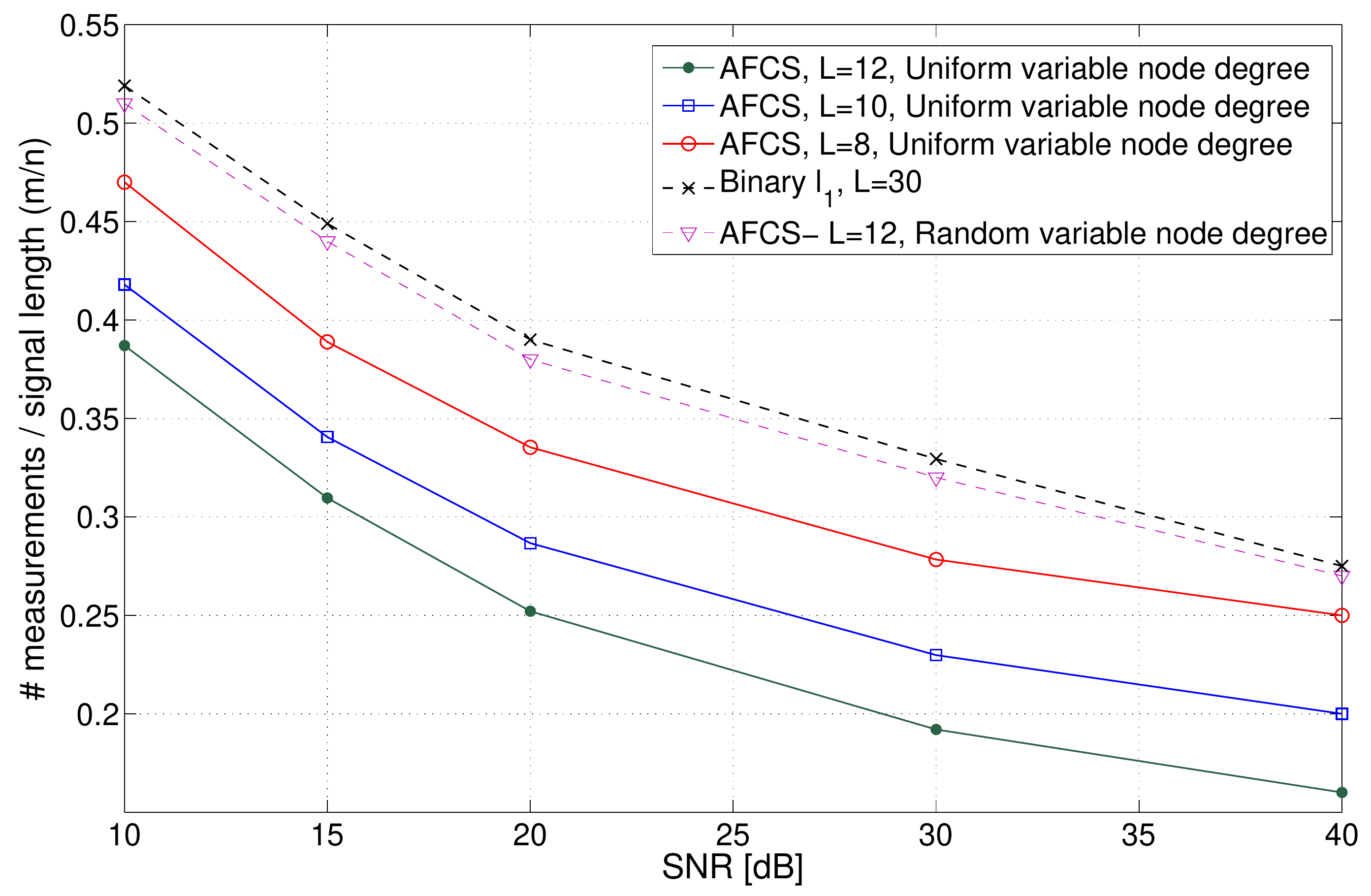}
\caption{Minimum sampling ratio required to successfully recover a sparse binary signal of length $n=1000$ versus the SNR, when the signal sparsity order is $s=0.1$.}
\label{MinNoisy}
\end{figure}

Fig. \ref{MinNoisy} shows the minimum sampling ratio ($\beta$) required for the successful reconstruction of all $k$-sparse binary signals ($k=100$) of length $n=1000$ versus the SNR. As can be seen in this figure, by increasing the measurement degree $L$ the performance of the AFCS scheme is improved. More specifically, when SNR = 30 dB, the AFCS scheme with $L=12$ requires 160 measurements, while it requires 200 and 240 measurements when $L=10$ and $L=8$, respectively. Note that the reconstruction algorithm proposed in \cite{BCS} only works in the noiseless settings. So, we only compare the proposed AFCS scheme with the $\ell_1$-minimization approach \cite{L1Binary} for the noisy case. As can be seen in Fig. \ref{MinNoisy}, the AFCS scheme with even a lower measurement degree always outperforms the $\ell_1$-minimization approach across all SNR values. The minimum sampling ratio for the proposed AFCS approach with non-uniform measurement matrix, where the variable node degree is not fixed, is also shown in Fig. \ref{MinNoisy}. As can be seen in this figure, the proposed AFCS approach with non-uniform variable node degrees and a very smaller measurement degree ($L=12$) performs very close to the binary $\ell_1$ scheme with a very large degree ($L=30$). This shows the superiority of the proposed scheme over conventional binary CS scheme for different measurement matrices and SNR values.

\section{Sparse Event Detection in WSNs using AFCS}
Recently, CS theory has been widely used in wireless sensor networks \cite{SenCompRec,AnEnEfficCS,EnSigAqusi,NonUnifCSWSN,CSforHarvest} and shown a great performance improvement in terms of network lifetime, energy efficiency and overall system throughput. Here, we consider the sparse event detection problem in WSNs, where a set of monitoring sensors try to capture a small number of active event sources amongst a large number of event resources, randomly placed inside the monitoring region. More specifically, we show how the proposed AFCS approach can be effectively used in WSNs for sparse event detection.
\subsection{Problem Definition}
\begin{figure}[t]
\centering
\includegraphics[scale=0.4]{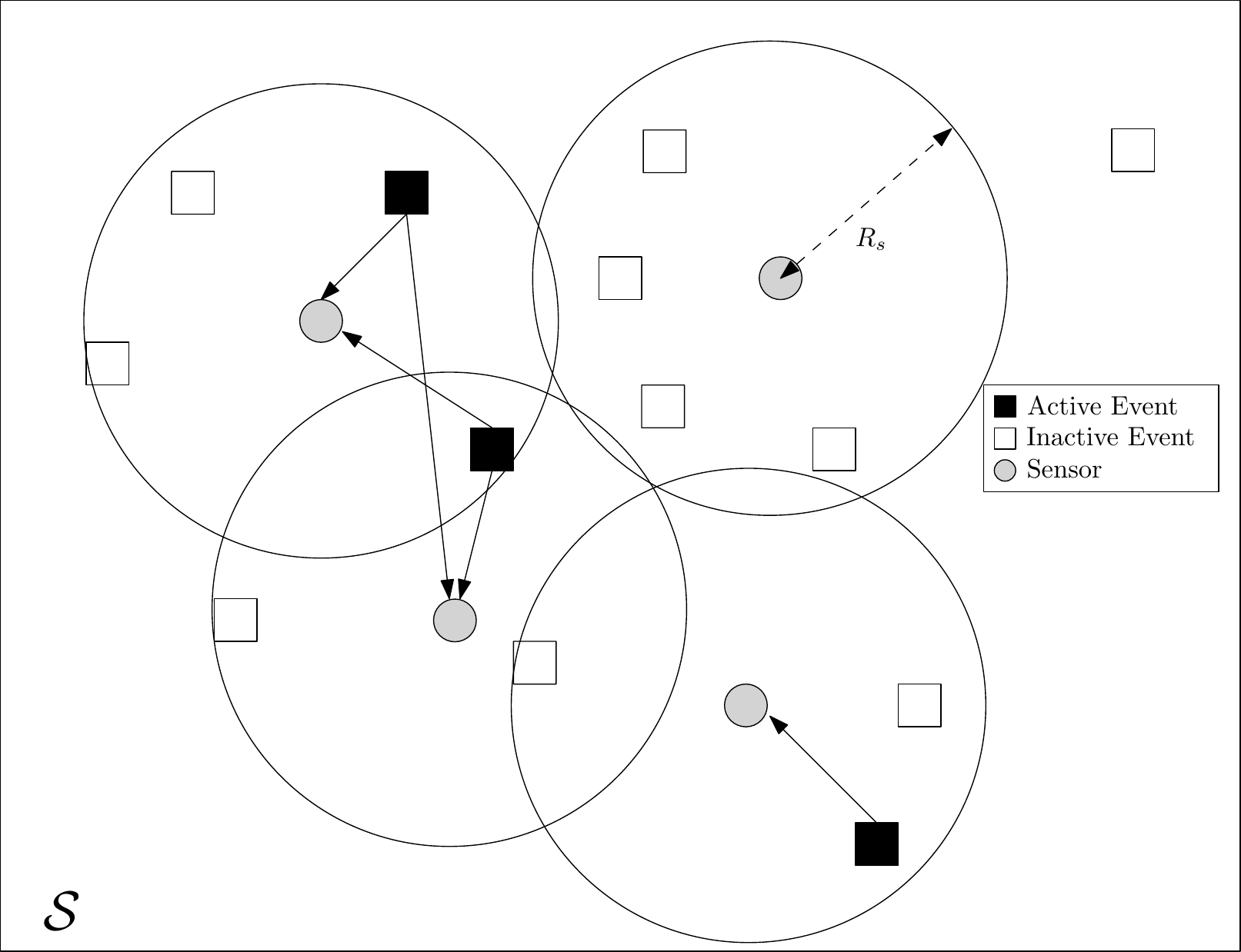}
\caption{$m$ active sensors are capturing $k$ out of $n$ active events.}
\label{wsnFig}
\end{figure}

We consider $n$ event sources, E$_1$, E$_2$,..., E$_n$, randomly placed in a monitoring region $\mathcal{S}$ with a total area of $S$ as depicted in Fig. \ref{wsnFig}. Each event source E$_i$ randomly generates an event $e_i$ to be measured by the sensor nodes. Here, an event is defined by a binary signal, which is either 0 or 1 (or a constant value). We refer to those events which have the value of 1 as active events, where the number of simultaneously active events, denoted by $k$, is much smaller than $n$, i.e., $k\ll n$ \cite{DecSpWSN}. Let $\textbf{e}_{n\times1}$ denote the event vector, where each of its elements, $e_i$ has a binary value, i.e., $e_i\in\{0,1\}$. Then, $\textbf{e}$ is a $k$-sparse binary signal of dimension $n$, i.e., $\sum_{i=1}^{n}e_i=k\ll n$.

The event sources are captured by $m$ active monitoring sensors, which are deployed in the monitoring region $\mathcal{S}$. The sensing range of each sensor is assumed to be $R_s$, which indicates that a sensor can receive the signal from event sources within a distance of at most $R_s$. In this paper, we consider two deployment scenarios. In the first scenario, referred to as \emph{random deployment}, sensors are randomly deployed in the monitoring region. In this approach, some event sources may not be covered by any sensors. As a result, these events cannot be detected. In the second scenario, referred to as \emph{uniform deployment}, sensors are uniformly distributed inside the monitoring region to fully cover the whole monitoring area. Let $h_{i,j}$ denote the channel between the $i^{th}$ event source and the $j^{th}$ sensor node, then the received signal $y_j$ at the $j^{th}$ sensor node can be written as follows:
\begin{align}
y_j=\sum_{i\in\mathcal{S}_j}h_{i,j}e_{i}+\epsilon_{j},
\end{align}
where  $\mathcal{S}_j$ is the set of event sources inside the sensing range of the $j^{th}$ sensor node, and $\epsilon_j$ is the thermal measurement noise at the $j^{th}$ sensor node. The received signal vector $\textbf{y}_{m\times 1}$ can then be written as follows:
\begin{align}
\label{problem}
\textbf{y}_{m\times 1}=\textbf{H}_{m\times n}\textbf{e}_{n\times1}+\epsilon_{m\times1},
\end{align}
where $\textbf{H}\triangleq[h_{i,j}]_{i,j}$ and $\textbf{e}=[e_i]_{1\le i\le n}$. 

The sink node is assumed to have full knowledge of the channel matrix \textbf{H} and all sensors' readings are sent to the sink node via a wireless channel at SNR $\gamma$ \cite{CSBSparseEvent}. More specifically, we assume the whole network is fully connected, which means that the sensor nodes can communicate with each other in the local communication range and each sensor node can correctly receive its neighbor's message \cite{CSBSparseEvent}. Generally, we assume that sensors' readings are relayed by the sensor nodes towards the sink node and they will be received at the sink node at SNR $\gamma$. The received signal at the sink node can then be shown as follows:
\begin{align}
\label{problem2}
\textbf{x}_{m\times 1}=\textbf{y}_{m\times 1}+\textbf{z}_{m\times1},
\end{align}
where $\textbf{z}$ is the $m\times1$ additive white Gaussian noise (AWGN) vector with zero mean and variance $\sigma_z^2$, and $\gamma=||\textbf{He}||^2/\sigma_z^2$. This model has been widely used for sparse event detection in WSNs \cite{ADLOwWSNS,BCSnew,DecSpWSN}. We will further discuss about the practical issues of this model in Section \ref{SecPrac} and propose some approach for its practical implementation.


\subsection{Sparse Event Detection in WSNs as an AFCS}
Since event sources generate binary signals, the event vector \textbf{e} can be considered as a vector of binary information symbols. As we assume that each sensor node receives signals only from active event sources inside the sensor's coverage area, the number of nonzero entries in each row of \textbf{H} is small. Thus the sensing process in (\ref{problem2}) is equivalent to the measurement process in AFCS, where the channel gains are considered as weight coefficients in the measurement matrix.  More specifically, the sensor degree is defined as the number of event sources inside the coverage area of a sensor node, determining the number of nonzero elements in the respective row of the measurement matrix. We further define the event degree as the number of sensor nodes which have covered that event source. The following lemma gives the probability distribution of the sensor degree in the uniform deployment scenario and gives a lower bound on the minimum number of sensor nodes required for having a minimum event degree of at least 1.
\begin{lemma}
\label{Degreelemma}
Let $n$ denote the number of event sources in the monitoring region $\mathcal{S}$, and let $m$ denote the number of sensor nodes with coverage radius $R_s$ uniformly placed inside $\mathcal{S}$ to detect active event sources. Let $\Omega_{u,d}$ denote the probability that the sensor degree is $d$. Then, $\Omega_{u,d}$ is given by:
\begin{align}
\label{degreeprob}
\Omega_{u,d}=\alpha_1f(n,d,P)+\alpha_2f(n,d,\frac{P}{4})+\alpha_3f(n,d,\frac{P}{2}),
\end{align}
where $P=\pi R_s^2/S $, $f(n,d,P)\triangleq\dbinom{n}{d}P^d(1-P)^{n-d}$, $\alpha_1=(m-4-4(\sqrt{m}-2))/m$, $\alpha_2=4/m$, and  $\alpha_3=(m-4(\sqrt{m}-2))/(m)$. Moreover, to guarantee that each event source is covered by at least one sensor node in the uniform deployment scenario, the following condition should be satisfied:
\begin{align}
\label{onecond}
m\ge\lceil\frac{\sqrt{S}}{2R_s}\rceil^2+\left(\lfloor\frac{\sqrt{S}}{2R_s}\rfloor+1\right)^2,
\end{align}
where $\lceil.\rceil$ and $\lfloor.\rfloor$ are the ceil and floor operators, respectively.
\end{lemma}

The proof of this lemma is provided in the Appendix \ref{proofdegree}. In the random deployment scenario, where both event sources and sensor nodes are randomly placed inside the sensing field, the sensor degree distribution $\Omega_{r,d}$ and event degree distribution $\Lambda_{r,d}$, defined as the probability mass function of the sensor degree and that of event degree, respectively, are given by:
\begin{align}
\nonumber
\Omega_{r,d}=\dbinom{m}{d}P^d(1-P)^{n-d},~~\Lambda_{r,d}=\dbinom{n}{d}P^d(1-P)^{m-d},
\end{align}
where $P=\pi R_s^2/S $. This arises from the fact that the probability that an event source is located inside the coverage area of a sensor node is the ratio of the coverage area of the sensor and area of the monitoring region. Since the locations of sensor nodes and event sources are independent, the number of event sources inside the coverage area of a sensor node is a binomial random variable. Similarly, we can show that the event degree follows a binomial distribution with parameter $m$ and success probability $P$.

\begin{figure}[t]
\centering
\includegraphics[scale=0.32]{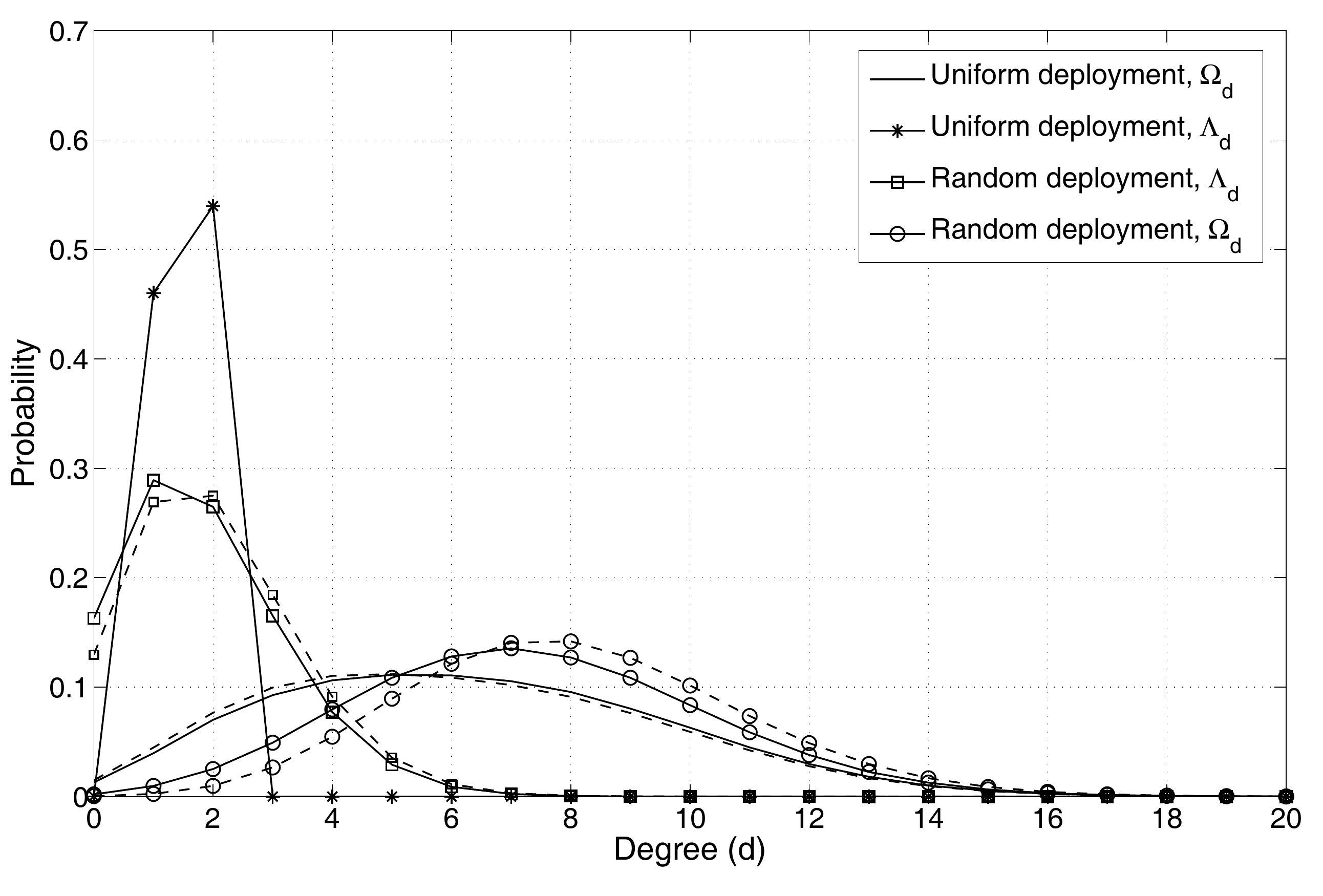}
\caption{The degree distribution of the sensor  and event sources for a WSN consisting of $m=64$ sensors with coverage radius $R_s=20~m$ to capture $n=256$ event sources within a rectangular field of dimension $200~m$ by $200~m$. Solid lines show simulation results and dashed lines show analytical ones.}
\label{Figdegrees}
\end{figure}

It is important to note that in the random deployment scenario, the probability that an event source is not covered by any sensor node is $\Lambda_{r,0}=(1-P)^m$. It is then straightforward to show that in the random deployment scenario, to guarantee that each event source is covered by at least one sensor node with the probability of at least $1-\zeta$ , i.e., $\Lambda_0\le\zeta$, the number of sensor nodes should satisfy the following condition:
\begin{align}
m\ge\frac{\ln(\zeta)}{\ln(1-P)},
\label{twocond}
\end{align}
where $\zeta>0$ is a given real number. Fig. \ref{Figdegrees} shows the sensor and event degree distribution in a WSN for both random and uniform deployment scenarios, where simulation and analytical results are very close. As can be seen in this figure, in the random deployment scenario the probability that an event source is not covered by any sensor node is relatively large (about 0.15); thus, those events cannot be detected during the event detection process. However, in the uniform deployment scenario, as long as condition (\ref{onecond}) is satisfied, each variable node is covered by at least one sensor node; therefore, the active events can be detected with a higher probability compared to the random deployment scenario.
\subsection{Analysis of the False Detection Probability}
As previously stated, the objective of the decoding algorithm is to find the conditional probabilities of the variable nodes for a given measurement matrix and the received signals. Since we consider the additive white Gaussian noise with variance $\sigma_z^2$, the conditional probabilities can be written as follows:
 \begin{align}
 \nonumber p(\textbf{e}|\textbf{x},\textbf{H})&=p(\textbf{z}=\textbf{x}-\textbf{He}|\textbf{H},\textbf{x})\\
 \nonumber &=\left(\frac{1}{\sqrt{2\pi \sigma_z^2}}\right)^m\prod_{j=1}^{m}\exp\left(-\frac{\left(x_j-\sum_{i=1}^{n}h_{j,i}e_i\right)^2}{2\sigma_z^2}\right),
 \end{align}
 and the decoding objective is to find $\textbf{e}$ in a way that $p(\textbf{e}|\textbf{x},\textbf{H})$ or equivalently $\log\left(p(\textbf{e}|\textbf{x},\textbf{H})\right)$ is maximized. Then, we have \cite{MahyarLetter}:
 \begin{align}
 \hat{\textbf{e}}=\arg \min_{\textbf{e}} ||\textbf{x}-\textbf{He}||_2^2.
 \end{align}
 The following lemma gives an upper bound on the probability of false detection (PFD), defined as the probability that a non-active event is falsely detected as active.
 \begin{lemma}
 \label{LemmaPE}
 Let $p_e$ denote the probability of false detection in a WSN consisting of $m$ sensor nodes with coverage radius $R_s$, which are placed in a monitoring region of area $S$. Then, $p_e$ is upper bounded as follows for a given noise variance $\sigma^2$:
 \begin{align}
 \label{uppeerror}
 p_e\le Q\left(\frac{1}{2\sigma_z}\sqrt{mP(1-P)R_s^{-\frac{3}{4}}}\right),
 \end{align}
 where $P=\pi R_s^2/S$ and $Q(x)=\frac{1}{\sqrt{2\pi}}\int_{x}^{\infty} e^{-\frac{x^2}{2}}dx$.
 \end{lemma}

The proof of this lemma is provided in Appendix \ref{prooflemmaPE}. It is important to note that from (\ref{uppeerror}) we can conclude that the probability of false detection in our proposed approach is very small. As an example, for a WSN consisting of $m=50$ sensors randomly placed in $500~m$ by $500~m$ monitoring area with coverage radius $R_s=50~m$, the probability of false detection is about $10^{-5}$ at $\gamma=20$ dB.

\section{Practical Considerations in the Sparse Event Detection in WSNs}
\label{SecPrac}
In this section, we address some practical issues in the system model, which have been commonly used for sparse event detection in WSNs. We modify the system model and consider more practical system setup to overcome these practical implementation issues. More specifically, we show that with the new system set up the problem is the same as the original problem defined in (\ref{problem}) and (\ref{problem2}), but with more practical assumptions.

Many existing schemes \cite{EventSNBinary,CSWSNLocal} applied CS techniques for event detection in wireless sensor networks. However, all of them directly apply the conventional compressive sensing algorithms. For compressive sensing to be applicable, they assume that the possible locations at which events may occur are known and the number of active events is also available. Furthermore, they also make the impractical assumption that each node is able to monitor every possible location. This essentially means that any sensor node covers the whole monitoring area, which is not true in reality. Therefore, these existing schemes are inapplicable for detection and localization of real-world events that may occur anywhere in the monitoring area, and cover limited region due to quick signal attenuation. More importantly, the number of occurred events is changing over time and unknown in advance.

\begin{figure}
  \centering
  \includegraphics[scale=0.56]{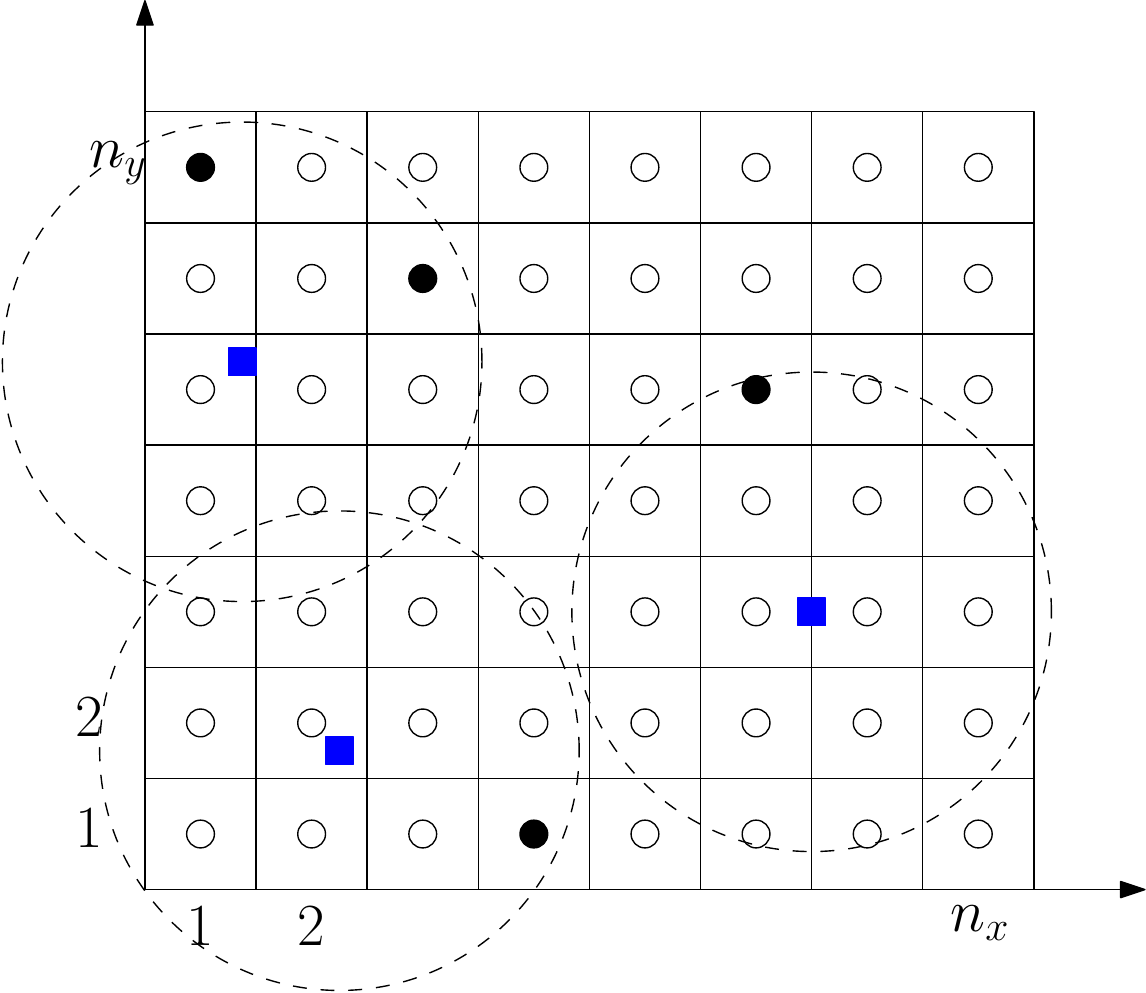}
  \caption{Event detection in WSNs using the grid approach. Squares show sensor nodes and black circles show active event sources.}
  \label{FigNewGrid}
\end{figure}

Unlike the conventional event detection approach for WSNs, we do not assume anything about the number of active events or their locations. We only assume that the value of each event source is either 0 or 1, where an active event has a value of 1, otherwise it is 0. As the sensor nodes do not know the location of the event sources, we assume that the monitoring region has an area of $S$ and is virtually divided into $n=n_x\times n_y$ small grids as shown in Fig. \ref{FigNewGrid}. Thus, the location of an event can be approximated by the centroid of the grid containing the event. Let $\mathcal{S}_j$ denote the set of grids inside the coverage area of sensor node $j$, then the received signal at sensor node $j$ can be shown as follows:
\begin{align}
y_j=\eta\sum_{i\in \mathcal{S}_j}\frac{e_i}{d^{\alpha/2}_{i,j}}+\epsilon_{j},
\end{align}
where $\epsilon_{j}$ is the thermal measurement noise at the $j^{th}$ sensor node, $d_{i,j}$ is the distance between the sensor node $j$ and the center of grid $i$, $\alpha$ is the propagation loss factor, $\eta$ is a fixed signal gain factor of sensor nodes, and $e_i=1$ if an active event is located in grid $i$, otherwise it is zero. This model has been widely used in WSNs \cite{MLCS,GCMULtiCS,EMCS,SparseTarget} and verified through practical experiments in \cite{CSWSNLocal} for the environment temperature. More specifically, the temperature 1 cm away from the heater was measured to be $60.0^o$ C, which is set as the source signal strength. The environmental temperature of the day conducting the experiments was about $30.4^o$ C. The experimental results in this paper have shown that the measured temperature at distance $d$ to the heater follows the following model:
\begin{align}
\nonumber t=30.4+\frac{60}{d^2},
\end{align}
which has been verified by the real measurements. This model can be further used when we assume that event sources are randomly distributed in the sensing field and event sources generate a signal of the same power $P_0$ when they are active. The measured signal power at distance $d$ of each event source can also be modeled by $P_d=P_0/d^{\alpha}$, where $\alpha$ is the path loss exponent. This model can be used in the environment that the wireless channel is modeled by path loss.

Sensor nodes transmit their observations along with their IDs to the sink node in a time divided manner, in which the total transmission time is divided into $m$ time slot and sensor $j$ transmits its observation in the $j^{th}$ time slot. The channel between sensor node $j$ and the sink node is known at the sink node which can be simply obtained through the periodical channel estimation at the sink node. Since the channel varies very slowly in sensor networks, the accurate channel state information can be acquired by pilot symbols in WSN \cite{CHCS}. It is important to note that as the sink node knows the location of each sensor node, it knows the grids which are inside the coverage area of each sensor node. Then, it can calculate the distance between the sensor nodes and the centre of each grid. We can also assume that each sensor node knows its location through embedded Global Positioning System (GPS) receivers or some localization algorithms \cite{CSWSNLocal}. Many other multi-hop transmission schemes can also be considered to deliver the measurements to the sink node, which are commonly used in WSNs. As these approaches have been widely studied in the literature and are out of scope of this paper, we refer the interested readers to \cite{CSWSNLocal} and references therein for further details.

We can then use the same system model as in (\ref{problem}), where $h_{i,j}=\eta/d_{i,j}^{\alpha/2}$ and $n$ is the number of grids. In fact, the sparse event detection can be modeled as a binary compressive sensing, where the number of measurements equals to the number of sensor nodes and the length of the binary sparse signal equals to the number of grids. It is then obvious that by increasing the number of grids, each active event falls inside one and only one grid. Since the number of active events is fixed (say $k$), by increasing the number of grids $n$, the binary vector $\textbf{e}$ becomes sparse.

We also make the practical assumption that the sensor network is roughly time synchronized. Note that we do not require accurate time synchronization. Many existing algorithms have been developed for practical time synchronization in large scale sensor networks. For example, FTSP \cite{CSWSNLocal} can achieve clock synchronization accuracy as much as 2.24 $\mu s$ by exchanging a few bytes among neighbors every 15 minutes. Since the event detection period is normally tens of seconds to tens of minutes. Thus, FTSP is sufficient for providing the needed time synchronization accuracy \cite{CSWSNLocal}.

It is very important to note that we assume that event sources are randomly distributed in the sensing field and only a small number of them are active. We also assume that each event source is active independently of other event sources. However, the measured signals at sensor nodes are correlated, as each sensor node has a sensing range which covers a small area of the sensing field which can overlap with other sensors' coverage areas. As we assume that the signal of interest is binary, i.e., each event source generates a signal with the known power when it is active, this correlation can be captured in the respective bipartite graph at the sink node. In fact, as the sink node knows the location of sensor nodes, which is a valid assumption and can be obtained through GPS data \cite{CSWSNLocal}, it also knows the coverage area of them, thus it can build a bipartite graph representing the connection between sensor nodes and event sources. If we assume that sensor nodes and the centroid of grids are shown respectively by check and variable nodes, then a variable node is connected to a check node if and only if the respective grid is located in the coverage area of the respective sensor node. In this way, sensor nodes which are closer to each other have more overlapped area, thus they have more common variable nodes (they are connected to a larger number of common variable nodes) in the bipartite graph. In other words, the correlation between sensing data can be well presented in the bipartite graph and has been already considered in the BP decoding and the respective analysis.

\section{Simulation Results}
In this section, we investigate the performance of the proposed AFCS scheme for sparse event detection in WSNs. More specifically, we consider two cases. First, a noiseless case is considered, where the sensor readings are noiseless, so the proposed sum verification decoder can be effectively used for the signal reconstruction from noiseless measurements. Second, we consider a general noisy case and simulate the performance of the AFCS scheme for different SNRs and signal sparsity orders, where the belief propagation decoder is used for the signal reconstruction. Here, the sampling ratio, $\beta$, is defined as the ratio of the number of sensors, $m$, and that of event sources, $n$. We use the probability of correct detection (PCD) as the performance metric, where PCD is defined as the percentage of correctly detected active events \cite{ADLOwWSNS}.
\begin{figure}[t]
\centering
\includegraphics[scale=0.34]{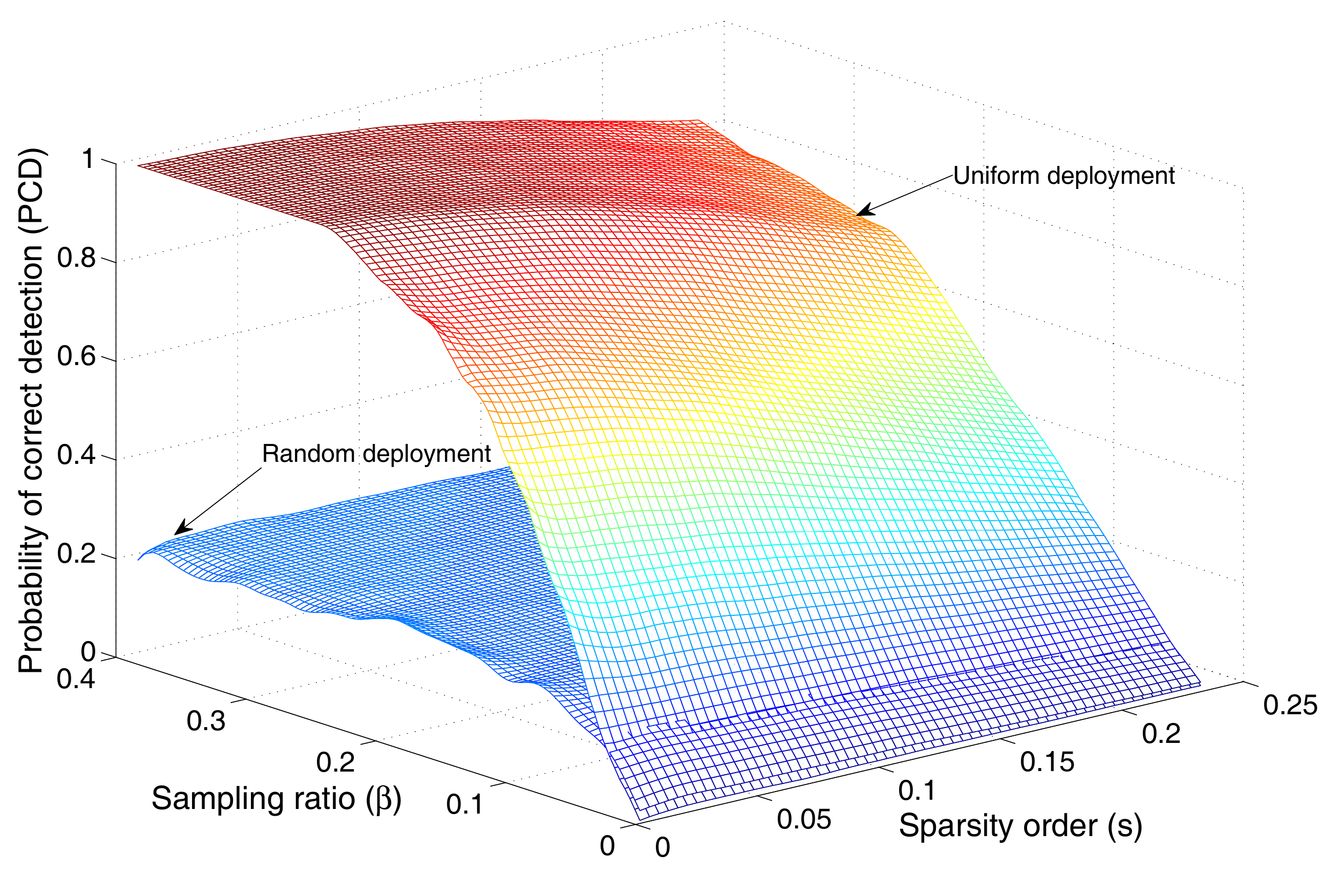}
\caption{Probability of correct detection versus the signal sparsity order and sampling ratio, when $R_s=50 m$ and $n=256$ event sources are randomly placed inside a $500~m$ by $500~m$ sensing field.}
\label{wsnsnoiseless}
\end{figure}

Fig. \ref{wsnsnoiseless} shows the probability of correct detection as a function of signal sparsity order and sampling ratio, when $R_s=50~m$ and the sum verification decoder is used with parameter $T=2$. As can be seen in this figure, the uniform deployment requires a less number of sensors to be deployed in the sensing field compared to the random deployment. This is because that in the random deployment scenario, the probability that an event source is not covered by any sensor node is relatively high even when the number of sensor nodes is large. Moreover, as shown in Fig. \ref{wsnsnoiseless}, when the sparsity order increases, the sum verification decoder requires more sensors' readings to detect all active events. The reason behind this is that when the number of active events increases, the number of active events connected to each sensor node also increases, thus the sum verification decoder cannot detect them. Moreover, when the sensors are randomly distributed, the degree of each event source follows the Poisson distribution, which results in a non-uniform error protection of the event sources. In other words, a fraction of event sources are covered by a large number of sensors, but the rest of event sources may be covered by a very small number of sensors; thus, they have a lower chance to be recovered from the measurements. It is important to note that the probability of false detection (PFD) is zero for the proposed sum verification decoder as it only detects the active events and other events are always detected as zero.

\begin{figure}[t]
\centering
\includegraphics[scale=0.34]{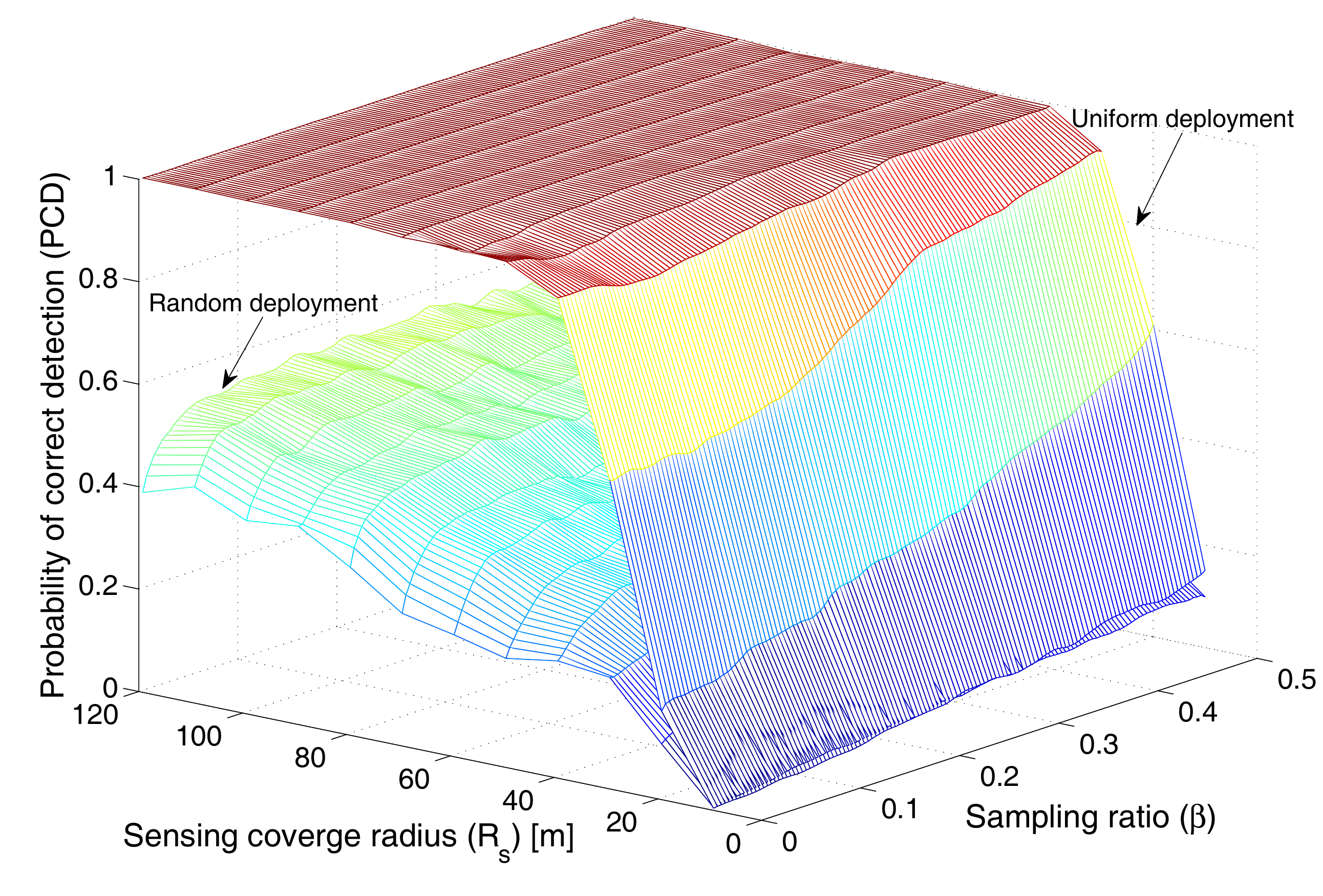}
\caption{Probability of correct detection versus the sensor coverage radius and sampling ratio, when only $k=10$ out of $n=256$ event sources are simultaneously active. Event sources are randomly placed inside a $500~m$ by $500~m$ sensing filed.}
\label{wsnsnoiselessRM}
\end{figure}

Fig. \ref{wsnsnoiselessRM} shows the PCD versus $R_s$ and the sampling ratio for the case that $n=256$ events sources are randomly distributed within a $500~m$ by $500~m$ sensing field, where only $k=10$ event sources are active. As can be seen in this figure, the proposed sum verification decoder can completely detect the active events in the uniform deployment scenario even at very small sampling ratios when the coverage radius of sensors is larger than $40 m$. In fact, by increasing the sensing coverage radius, active event sources are covered by more sensor nodes and thus they can be detected with a higher probability. However, as can be seen in this figure, in the random deployment scenario active event sources cannot be complectly detected even with a large sensing coverage radius or high sampling ratio. 

\begin{figure}[t]
\centering
\includegraphics[scale=0.34]{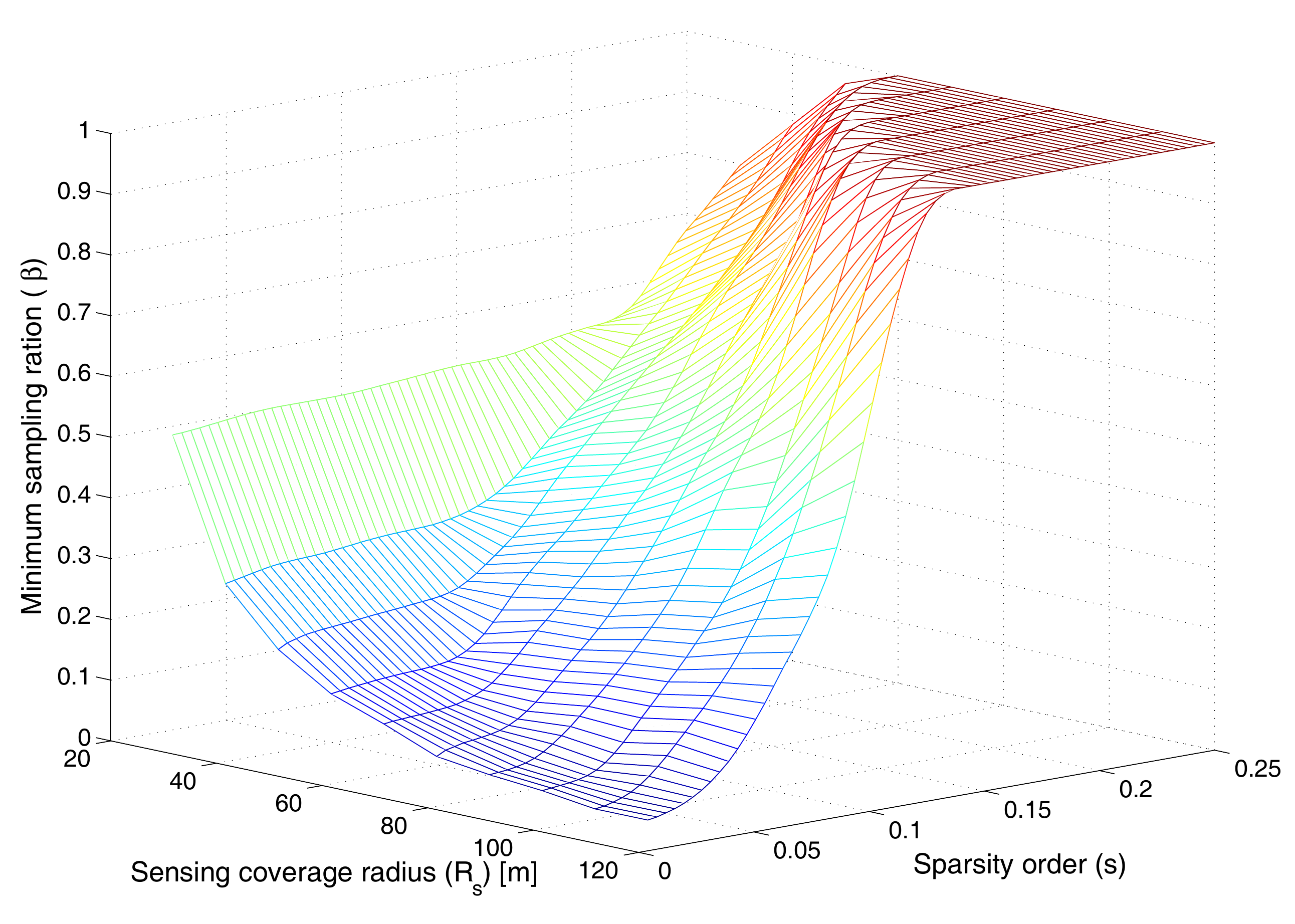}
\caption{Minimum sampling ratio ($\beta$) required to successfully detect active events amongst $n=256$ event sources randomly distributed within an area of $500~m$ by $500~m$ versus $R_s$ and $s$. The sensor nodes are uniformly placed inside the sensing field.}
\label{optmbeta}
\end{figure}

Fig. \ref{optmbeta} shows the minimum sampling ratio required to successfully detect all active events versus the signal sparsity order $s$ and $R_s$ for the uniform deployment scenario. As can be seen in this figure, with increasing $s$, a larger number of sensors have to be deployed in the sensing field. This is because of the fact that a larger number of active event sources are connected to each sensor nodes; thus, the proposed sum verification decoder with $T=2$ requires a larger number of measurements for sparse event detection. Moreover, when $R_s$ increases, a smaller number of sensor nodes is required to be deployed in the sensing field to successfully recover all active events. However, when the signal sparsity order increases, a lower number of sensor nodes is required when the sensor coverage radius is low compared to the case that the sensing coverage radius is high.

\begin{figure}[t]
\centering
\includegraphics[scale=0.35]{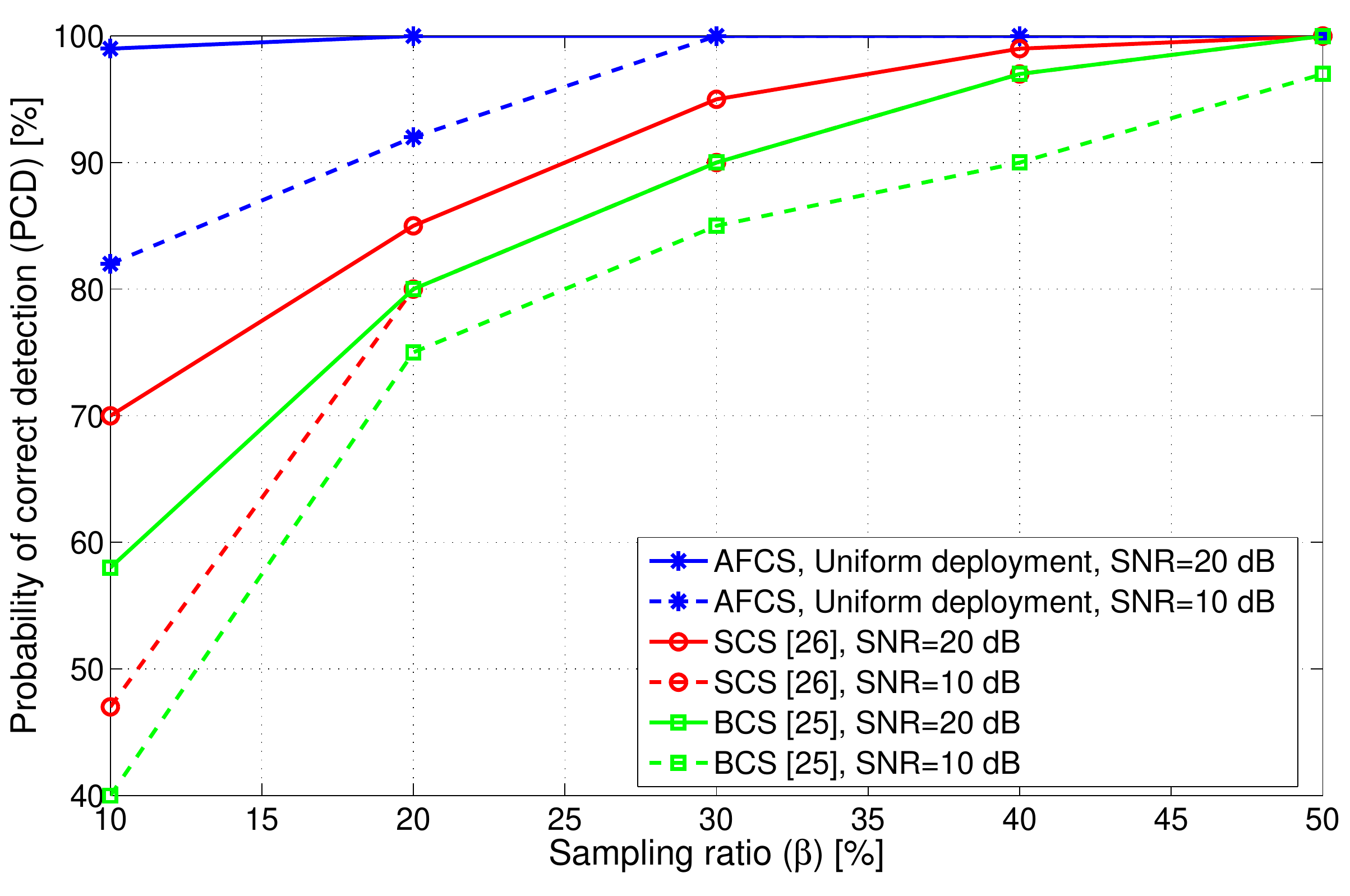}
\caption{Probability of correct detection (PCD) versus the sampling ratio ($\beta$) at different SNRs, when $n=1000$ event sources are randomly distributed in an area of $1000~m$ by $1000~m$ and $k=10$ events are simultaneously active.}
\label{PCDfig}
\end{figure}

We now compare the performance of AFCS with Bayesian compressive sensing (BCS) \cite{BCSnew} and sequential compressive sensing (SCS) \cite{ADLOwWSNS} at different SNRs and sampling ratios. Fig. \ref{PCDfig} shows PCD versus the sampling ratio at different SNR values, when $n=1000$ event sources are randomly distributed in an area of $1000~m$ by $1000~m$, $k=10$, and $\alpha=3$. As can be seen in this figure, the proposed AFCS scheme outperforms both BCS \cite{BCSnew} and SCS \cite{ADLOwWSNS} in terms of PCD for different SNRs. Moreover, the number of required sensor nodes in SCS is relatively large (500), which significantly increases the total number of transmissions, leading to poor energy efficiency. It is important to note that the existing approaches need to know the number of active events to achieve a relatively good performance in terms of PCD; however, the proposed AFCS approach does not need to know the sparsity level of the events and can work for various numbers of active events. Moreover, the achievable PCD for the AFCS scheme with random deployment is very close to that for the SCS approach \cite{ADLOwWSNS}, and so we removed the respective curve from the figure for a clearer representation.

\begin{figure}[t]
\centering
\includegraphics[scale=0.34]{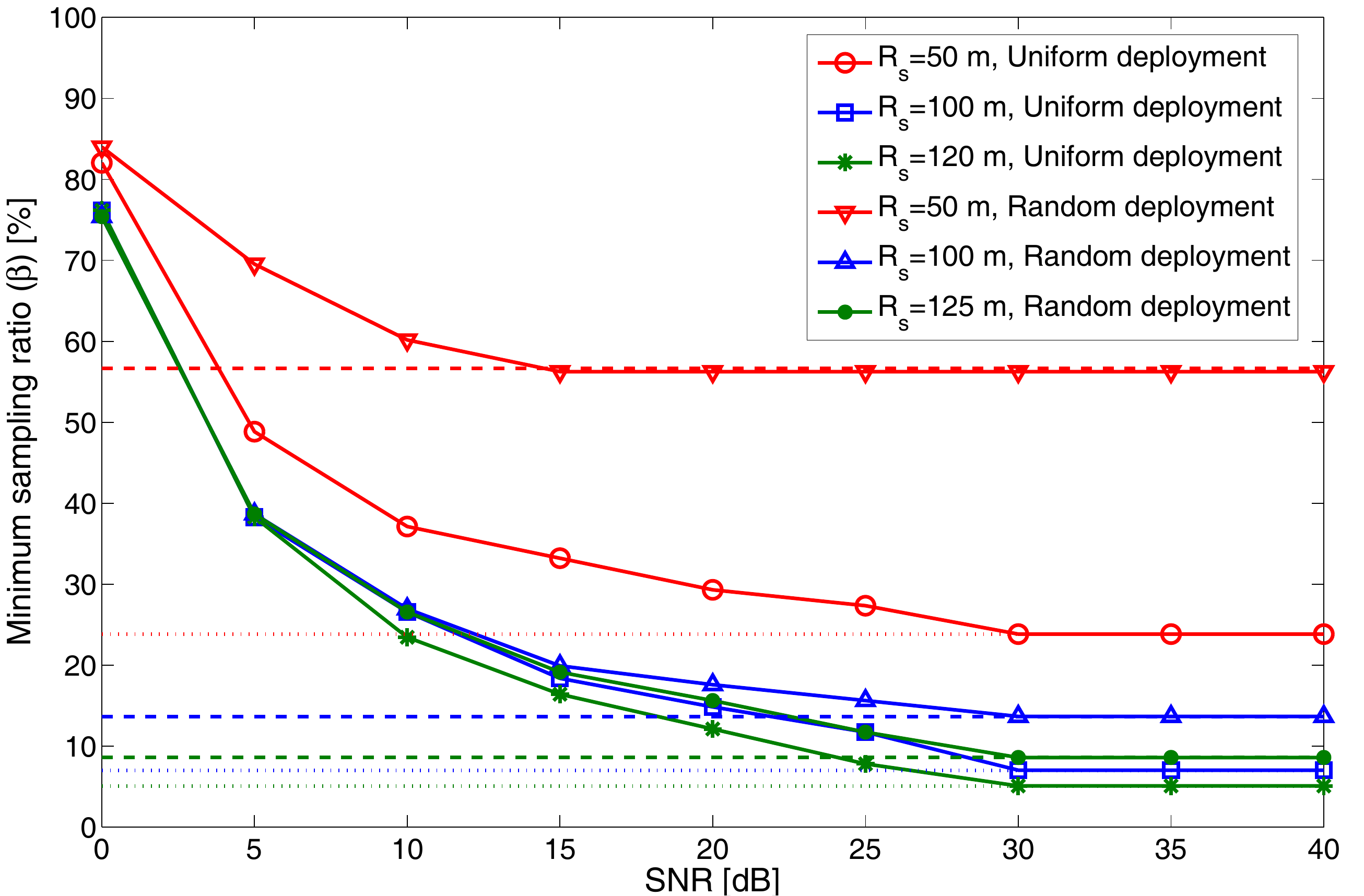}
\caption{Minimum sampling ratio for the proposed AFCS scheme to achieve a PCD higher than $99\%$, when $n=256$ event sources randomly distributed with the 500m-by-500m area and only $k=4$ events are active. Dashed line show the lower bound on the minimum number of sensors obtained from (\ref{onecond}) and (\ref{twocond}).}
\label{MinPCD}
\end{figure}

Fig. \ref{MinPCD} shows the minimum sampling ratio versus SNR in order to achieve a PCD larger than or equal to $99\%$, when $\alpha=3$ and $n=256$ event sources are randomly distributed within $500~m$ by $500~m$ area. As can be seen in this figure, in the uniform and random deployment scenarios, the minimum number of nodes approaches the lower bounds in (\ref{onecond}) and (\ref{twocond}), when SNR increases. Moreover, in low SNRs, the proposed approach can detect the active events with a large number of sensor nodes, where a WSN with sensors with a high coverage radius is more robust to the noise effect in terms of the sampling ratio. Note that in the random deployment scenario, more sensor nodes are required as some event sources may not be covered by any sensor nodes, leading to a lower percentage of correct detection.

\section{Conclusion}
In this paper, we proposed an analog fountain compressive sensing (AFCS) approach for sparse recovery of the binary sparse signal. In the proposed approach, each measurement is generated from the binary sparse signal in a way that the variable node degrees are almost the same. We proposed a generalized verification based reconstruction algorithm, which enables a measurement connected to at most $T$ nonzero signal elements to recover their connected variable nodes. The measurement degree was also optimized to maximize the number of the verified variable nodes by each measurement. We showed that the number of measurements required for the successful recovery of the sparse signal is of $\mathcal{O}(-n\log(1-s))$, where $n$ is the signal length and $s$ is the sparsity order. Simulation results showed that the AFCS scheme significantly outperforms the conventional binary CS and $\ell_1$-minimization approaches in terms of the error rate for different sparsity orders. We further showed that the sparse event detection problem in wireless sensor networks can be represented by AFCS. Simulation results showed that the proposed AFCS approach outperforms the existing sparse event detection schemes in WSNs in terms of the probability of correct detection in a wide range of signal to noise ratios with a much smaller number of sensor nodes and a negligible probability of false detection.

\appendices

\section{Proof of Lemma \ref{sumortreelemma}}
\label{proofsumor}
Let $f^{(0)}_i$ and $f^{(1)}_i$ denote the probability that a check node can verify its connected variable node as zero and nonzero variable node, respectively, after $i$ iterations of the decoding algorithm. Let us define $f_{i}\triangleq f^{(0)}_i+f^{(1)}_i$, which denotes the probability that a check node can verify the connected variable nodes in the $i^{th}$ iteration of the sum verification decoding. It is clear that a SUM node at depth $2i-1$ can verify its parent at depth $2(i-1)$ as zero variable node, if one of the following events happened. First, all of its connected OR nodes at depth $2i$ have been verified; second, at least one of its connected OR nodes at depth $2i$ has not been verified, but the SUM node can verify the remaining variable nodes. The second event happens if the value of the check node is equal to the sum of at most $T$ edges connected to the unverified variable nodes. Thus,  $f^{(0)}_i$ can be written as follows:
\begin{align}
\label{verzero}
f^{(0)}_i=\sum_{i=0}^{d}\sum_{j=0}^{A_1}\dbinom{d}{i}\dbinom{d-i}{j}p_{i-1}^i\left(q_{i-1}^{(0)}(1-p_{i-1})\right)^{d-i}\lambda_{i-1}^j,
\end{align}
which arises from the fact that $i$ out of $d$ variable nodes connected to the check node are not verified with probability $\dbinom{d}{i}p_{i-1}^i(1-p_{i-1})^{d-i}$, and $j$ out of $d-i$ unverified variable nodes have potentially nonzero values with probability $\dbinom{d-i}{j}(q_{i-1}^{(0)})^{d-i-j}(q_{i-1}^{(1)})^{j}$. Since the probability that an unverified variable node at depth $2(i-1)$ being zero is $q_{i-1}^{(0)}$, and $i$ and $j$ vary from $0$ to $d$ and $0$ to $A_1\triangleq \min\{T-1,d-i\}$, respectively, then (\ref{verzero}) is straightforward. Similarly, $f^{(1)}_i$ can be calculated as follows:
\begin{align}
\label{vernonzero}
f^{(1)}_i=\sum_{i=0}^{d}\sum_{j=0}^{A_2}\dbinom{d}{i}\dbinom{d-i}{j}p_{i-1}^i\left(q_{i-1}^{(0)}(1-p_{i-1})\right)^{d-i}\lambda_{i-1}^{j+1},
\end{align}
where $A_2=\min\{T-1,d-i\}$. As we consider that $f_i= f^{(0)}_i+f^{(1)}_i$, then (\ref{lemmapc}) can be directly obtained by the summation of (\ref{verzero}) and (\ref{vernonzero}). Moreover, if a variable node receives at least one message of value $1$ form its connected check nodes, then it can be verified. Let us assume that an OR node has $j$ children with probability $\delta_j$, so it can be verified with probability $1-\delta_j(1-f_i)^j$. Then, $p_i$ can be verified by summation over $j$ as follows:
\begin{align}
p_i=\sum_{j}(1-\delta_j(1-f_i)^j)=1-\delta(1-f_i).
\end{align}
As stated before, after $i$ iterations of the verification-based decoding algorithm, a variable node can be verified as zero variable node with probability $f^{(0)}_i$. This means that the average number of zero valued unverified variable nodes is $k-nf^{(1)}_i$. Since the average number of unverified variable nodes in the $i^{th}$ iterations is $n-np_i$, then $q_{i}^{(1)}$ can be calculated as follows:
\begin{align}
q_{i}^{(1)}=\frac{k-nf^{(1)}_i}{n-p_i}=\frac{1-q_{0}^{(0)}-f^{(1)}_i}{1-p_i}.
\end{align}
This completes the proof.
\section{Proof of Lemma \ref{OptLlemma}}
\label{proofOptLemma}
For simplicity, we assume $L\triangleq d+1$, then by using the fact that $\dbinom{L}{i}$ can be approximated by $L^i/i!$ for $L$ being very large compared to $i$, the average number of recovered variable nodes by each measurement, denoted by $R$, can be approximated as follows for a small $T$:
\begin{align}
\label{sapprox}
R\approx\sum_{i=0}^{T}\frac{L^{i+1}}{i!}q_0^{L-i}(1-q_0)^i,
\end{align}
and the derivative of $R$ with respect to $L$ can be calculated as follows:
\begin{align}
\frac{d}{dL}(R)=p^L\sum_{i=1}^{T}\frac{\lambda^i}{i!}L^i\left((i+1)+L\log(q_0)\right),
\end{align}
where $\lambda\triangleq(1-q_0)/q_0$. The solution for the equation $dR/dL=0$  gives the optimal value of $L$. Therefore, we need to solve the following equation:
\begin{align}
\sum_{i=1}^{T}\frac{\lambda^i}{i!}L^i\left((i+1)+L\log(q_0)\right)=0.
\end{align}
The term inside the summation is zero, when $L=\frac{i+1}{-\log(q_0)}$, which gives us an approximation of the optimum $L$. A better approximation can be achieved by taking the average of $\frac{i+1}{-\log(q_0)}$ over $i$, which directly results in (\ref{Loptlemma}), due to the fact that $\sum_{i=1}^{T+1}i=(T+1)(T+2)/2$.

\section{Proof of Lemma \ref{orderlemma}}
\label{prooforder}
It is clear that each measurement can verify at most $L_{opt}$ variable nodes; thus, the destination requires at least $n/L_{opt}$ measurements to  determine all variable nodes. Moreover, each measurement can verify its connected variable nodes with probability $P(L_{opt})$, and accordingly the average number of measurements that can recover their variable nodes is $mP(L_{opt})$. However, after the first iteration of the verification-based decoding algorithm, and removing the edges connected to the verified variable nodes, a new set of measurements can recover their variable nodes in the next iteration of the reconstruction algorithm and so, the verification process is continued. This means that the actual number of required measurements is less than $n/mP(L_{opt})$. Therefore, we have:
\begin{align}
\label{ineqMopt}
\frac{n}{L_{opt}}\le m\le \frac{n}{L_{opt}p(L_{opt})},
\end{align}
where
\begin{align}
\label{plopt}
p(L_{opt})=\sum_{j=0}^{T}\dbinom{L_{opt}}{j}(1-s)^{L_{opt}-j}s^j.
\end{align}
As can been seen in (\ref{plopt}), $p(L_{opt})$ is an increasing function of $T$. Thus, we have
\begin{align}
P(L_{opt})>\sum_{j=0}^{0}\dbinom{L_{opt}}{j}(1-s)^{L_{opt}-j}s^j=(1-s)^{L_{opt}},
\end{align}
and by replacing (\ref{Loptlemma}) in the above equation for $T=0$, we have
\begin{align}
P(L_{opt})>(1-s)^{\frac{-1}{\log(1-s)}}=e^{-1}.
\end{align}
Therefore, (\ref{ineqMopt}) can be rewritten as follows:
\begin{align}
\label{optm2}
\frac{n}{L_{opt}}\le m\le \frac{ne}{L_{opt}}
\end{align}
and by replacing (\ref{Loptlemma}) in (\ref{optm2}), we have:
 \begin{align}
\frac{-2n\log(1-s)}{T+2}\le m\le \frac{-2ne\log(1-s)}{T+2}.
\end{align}
This completes the proof.

\section{Proof of Lemma \ref{Degreelemma}}
\label{proofdegree}
\begin{figure}[t]
\centering
\includegraphics[scale=0.6]{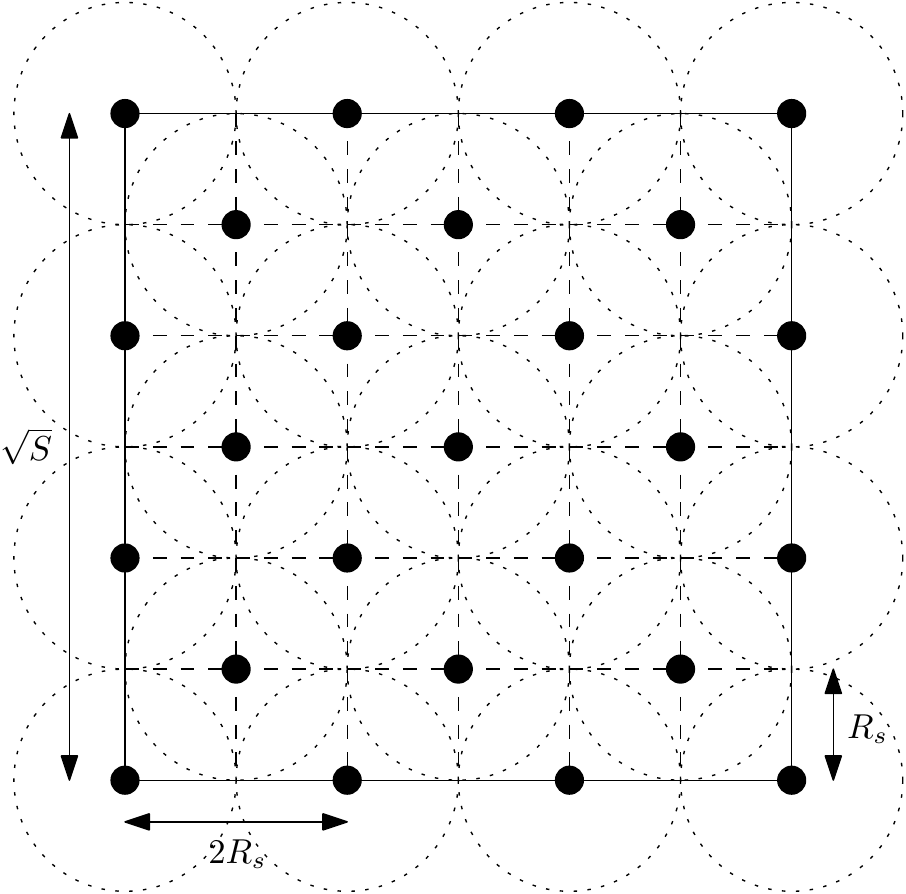}
\caption{Uniform deployment of sensor nodes in  the sensing field.}
\label{graphproof}
\end{figure}

As we assume that event sources are randomly placed in the sensing field $\mathcal{S}$, the probability that an event source is located inside the coverage area of a sensor node with a total coverage area of $\pi R_s^2$ will be $P=\pi R_s^2/S$. Moreover, as event sources are independently distributed in the monitoring field, the probability that $d$ out of $n$ event sources exist in the coverage area of a sensor node will be $f(n,d,P)$. However, as shown in Fig. \ref{graphproof}, four sensor nodes, which are located at the edges, will have a total coverage area of $\pi R_s^2/4$ inside $\mathcal{S}$ and $4(\sqrt{m}-2)$ sensor nodes have coverage area of $\pi R_s^2/2$ inside $\mathcal{S}$. Thus, according to the law of total probability, the probability that $d$ out of $n$ event sources are located inside the coverage area of a sensor node is given by (\ref{degreeprob}).

As can be clearly seen in Fig. \ref{graphproof}, the maximum distance between two neighboring sensor nodes is $2R_s$; thus, the minimum number of sensor nodes required to cover the sensing field in a way that sensors coverage areas do not intersect, is $\lceil\frac{\sqrt{S}}{2R_s}\rceil^2$. However, to cover the rest of the sensing field, we need at least  $\left(\lfloor\frac{\sqrt{S}}{2R_s}\rfloor+1\right)^2$ sensor nodes. Therefore, the minimum number of sensor nodes to fully cover the sensing field is given by (\ref{onecond}).

\section{Proof of Lemma \ref{LemmaPE}}
\label{prooflemmaPE}
Let us consider that signal $\textbf{e}'$ has been decoded, with $\ell$ false detections. Without loss of generality, we consider that $\textbf{e}$ and $\textbf{e}'$ are different in the first $\ell$ places, and let $p_{\ell}$ denote the probability of this event, then we have,
 \begin{align}
 \nonumber p_{\ell}&=p\left(||\textbf{x}-\textbf{He}'||_2^2\le ||\textbf{x}-\textbf{He}||_2^2\right)\\
 \nonumber &=p\left(\sum_{j=1}^{m}\left(x_j-\sum_{i=1}^{n}h_{j,i}e'_i\right)^2\le\sum_{j=1}^{m}\left(x_j-\sum_{i=1}^{n}h_{j,i}e_i\right)^2\right)\\
 \nonumber &\overset{(a)}=p\left(\sum_{j=1}^{m}\left(z_j+\sum_{i=1}^{n}h_{j,i}(e_i-e'_i)\right)^2\le\sum_{j=1}^{m}z_j^2\right)\\
 \nonumber &\overset{(b)}=p\left(\sum_{j=1}^{m}\left(z_j+\sum_{i=1}^{\ell}h_{j,i}(e_i-e'_i)\right)^2\le\sum_{j=1}^{m}z_j^2\right)\\
 \nonumber &=p\left(\sum_{j=1}^{m}\left(\sum_{i=1}^{n}h_{j,i}(e_i-e'_i)\right)^2\le\sum_{j=1}^{m}\sum_{i=1}^{n}2z_jh_{j,i}(e'_i-e_i)\right),
 \end{align}
 where step $(a)$ follows the fact that $\textbf{x}=\textbf{He}+\textbf{z}$, and step $(b)$ follows the fact that only first  $\ell$ places of $\textbf{e}$ and $\textbf{e}'$ are different. As noise components $w_j$ are i.i.d. Gaussian random variables with variance $\sigma_z^2$, then $\sum_{j=1}^{m}\sum_{i=1}^{\ell}2z_jh_{j,i}(e'_i-e_i)$ also has a zero mean Gaussian random variable with variance $4\sigma_z^2\sum_{j=1}^{m}\left(\sum_{i=1}^{\ell}h_{j,i}(e_i-e'_i)\right)^2$. Thus, $p_{\ell}$ can be calculated as follows:
 \begin{align}
 \label{errorML}
 p_{\ell}=Q\left(\frac{1}{2\sigma_z}\sqrt{\sum_{j=1}^{m}\left(\sum_{i=1}^{\ell}h_{j,i}(e_i-e'_i)\right)^2}\right).
 \end{align}
 Let $c_i\triangleq e_i-e'_i$, then $c_i=-1$ when a false detection occurs. 
It is then clear that $p_{\ell}$ is a decreasing function of $\ell$; thus, the probability of false detection, $p_e$, which is a linear summation of $p_{\ell}$'s is always smaller or equal to $p_1$, i.e., $p_e\le p_1$. Thus, we have:
\begin{align}
\nonumber p_e\le Q\left(\frac{1}{2\sigma_z}\sqrt{\sum_{j=1}^{m}h_{j,1}^2}\right).
\end{align}
The event degree follows a binomial distribution of parameter $m$ and success probability $P=\pi R_s^2/S$  with average event degree of $mP(1-P)$, and $\mathbb{E}[h_{j,i}^2]\le R_s^{-\frac{3}{4}}$; thus, (\ref{uppeerror}) is straightforward.

\bibliographystyle{IEEEtran}
\footnotesize
\bibliography{IEEEabrv,sample2}

\end{document}